\documentclass[%
 reprint,
superscriptaddress,
nofootinbib,
 amsmath,amssymb,
 aps,
]{revtex4-1}

\usepackage{graphicx}
\usepackage{bm}
\usepackage{xcolor}
\usepackage{array}
\usepackage{arydshln}
\usepackage{makecell}
\usepackage{multirow}
\usepackage{colortbl}

\usepackage{hyperref}

\definecolor{MyBlue}{rgb}{0.15,0.15,0.70}
\hypersetup{
colorlinks=true,
citecolor=MyBlue,
linkcolor=MyBlue,
urlcolor=MyBlue
}

\newcommand{\mmin}{m_{\rm min}}
\newcommand{\mmax}{m_{\rm max}}

\newcommand{\dgw}{d_L^{\,\rm gw}}
\newcommand{\dem}{d_L^{\,\rm em}}
\newcommand{\dcom}{d_{\rm com}}

\newcommand{\nn}{\nonumber}

\renewcommand\({\left(}
\renewcommand\){\right)}
\renewcommand\[{\left[}
\renewcommand\]{\right]}

\newcommand{\ra}{\rightarrow}

\def\lsim{\raise 0.4ex\hbox{$<$}\kern -0.8em\lower 0.62
ex\hbox{$\sim$}}

\def\gsim{\raise 0.4ex\hbox{$>$}\kern -0.7em\lower 0.62
ex\hbox{$\sim$}}

\def\lbar{{\hbox{$\lambda$}\kern -0.7em\raise 0.6ex
\hbox{$-$}}}

\newcommand\eq[1]{eq.~(\ref{#1})}
\newcommand\eqs[2]{eqs.~(\ref{#1}) and (\ref{#2})}

\newcommand\p{\partial}

\newcommand\ee{\end{equation}}
\newcommand\be{\begin{equation}}
\def\bea{\begin{array}}
\def\eea{\end{array}}\def\ea{\end{array}}
\newcommand\ees{\end{eqnarray}}
\newcommand\bees{\begin{eqnarray}}
\def\nn{\nonumber}

\def\dslash{\hspace{-1mm}\not{\hbox{\kern-2pt $\partial$}}}
\def\Dslash{\not{\hbox{\kern-2pt $D$}}}
\def\pslash{\not{\hbox{\kern-2.1pt $p$}}}
\def\kslash{\not{\hbox{\kern-2.3pt $k$}}}
\def\qslash{\not{\hbox{\kern-2.3pt $q$}}}

\def\p1{{\bf p}_1}
\def\p2{{\bf p}_2}
\def\k1{{\bf k}_1}
\def\k2{{\bf k}_2}

\newcommand{\dddM}{\kern 0.2em \raise 1.9ex\hbox{$...$}\kern -1.0em \hbox{$M$}}
\newcommand{\dddQ}{\kern 0.2em \raise 1.9ex\hbox{$...$}\kern -1.0em \hbox{$Q$}}
\newcommand{\dddI}{\kern 0.2em \raise 1.9ex\hbox{$...$}\kern -1.0em\hbox{$I$}}
\newcommand{\dddJ}{\kern 0.2em \raise 1.9ex\hbox{$...$}\kern-1.0em
\hbox{$J$}}
\newcommand{\dddcalJ}{\kern 0.2em \raise 1.9ex\hbox{$...$}\kern-1.0em
\hbox{${\cal J}$}}

\newcommand{\dddO}{\kern 0.2em \raise 1.9ex\hbox{$...$}\kern -1.0em
\hbox{${\cal O}$}}
\def\dddz{\raise 1.5ex\hbox{$...$}\kern -0.8em \hbox{$z$}}
\def\dddd{\raise 1.8ex\hbox{$...$}\kern -0.8em \hbox{$d$}}
\def\dddbd{\raise 1.8ex\hbox{$...$}\kern -0.8em \hbox{${\bf d}$}}
\def\ddbd{\raise 1.8ex\hbox{$..$}\kern -0.8em \hbox{${\bf d}$}}
\def\dddx{\raise 1.6ex\hbox{$...$}\kern -0.8em \hbox{$x$}}

\newcommand{\msun}{M_{\odot}}

\newcommand{\ode}{\Omega_{\rm DE}}
\newcommand{\oma}{\Omega_{M}}

\newcommand{\rde}{\rho_{\rm DE}}

\begin{document}

\preprint{APS/123-QED}

\title{Cosmology and modified gravitational wave propagation \\from  binary black hole population models}

\author{Michele Mancarella}
\email{michele.mancarella@unige.ch}
\affiliation{D\'epartement de Physique Th\'eorique and Center for Astroparticle Physics,\\
Universit\'e de Gen\`eve, 24 quai Ansermet, CH--1211 Gen\`eve 4, Switzerland}
\author{Edwin Genoud-Prachex}%
\email{genoud@itp.uni-frankfurt.de}
\affiliation{Institute for Theoretical Physics, Goethe University, 60438 Frankfurt am Main, Germany}
\author{Michele Maggiore}
\email{michele.maggiore@unige.ch}
\affiliation{D\'epartement de Physique Th\'eorique and Center for Astroparticle Physics,\\
Universit\'e de Gen\`eve, 24 quai Ansermet, CH--1211 Gen\`eve 4, Switzerland}

\date{\today}

\begin{abstract}
A joint hierarchical Bayesian analysis of the binary black hole (BBH) mass function, merger rate evolution and cosmological parameters can be used to extract information on both the cosmological and population parameters. We extend this technique to include the effect of modified gravitational wave (GW) propagation. We discuss the constraints on the parameter $\Xi_0$ that describes  this phenomenon (with $\Xi_0=1$ in General Relativity, GR) using the data from the GWTC-3 catalog. We find  the  constraints $\Xi_0 = 1.2^{+0.7}_{-0.7}$ with a flat prior on $\Xi_0$, and $\Xi_0 = 1.0^{+0.4}_{-0.8}$ with a prior uniform in $\log\Xi_0$ ($68\%$ C.L., maximum posterior and HDI), which only rely on the presence of a feature in the BBH mass distribution  around $\sim 30-45 \msun$, and are robust to whether or not the event GW190521 is considered an outlier of the population. 
We then study in more detail the effects of modified GW propagation on population and cosmological analyses for LIGO/Virgo at design sensitivity. We find that, for a given data-taking period, the relative error $\Delta\Xi_0/\Xi_0$ has a significant dependence on the fiducial value of $\Xi_0$, since the latter has a strong influence on the detection rate. For five years of data, the accuracy ranges from $\sim 10\%$ on $\Xi_0$ when $\Xi_0=1$ to
$\Delta\Xi_0/\Xi_0\sim  20\%$  for $\Xi_0=1.8$, that represents a large deviation from GR, still consistent with current limits and predicted by viable cosmological models. For the Hubble parameter,
we  forecast an accuracy of $\Delta H_0/H_0 \sim 20\%$, and an accuracy on $H(z)$ of $\sim7\%$ at a pivot redshift $z_*\sim 0.8$; this updates the results  found  in previous studies, by making use of a population model compatible with current observations, and is  about a factor 2 worse, in twice the observing time.
We finally show that, if Nature is described by a modified gravity theory with a large deviation from the GR value  $\Xi_0=1$, such as $\Xi_0=1.8$,  analysing the data assuming GR produces a significant bias in the inferred values of the mass scales, Hubble constant, and particularly the BBH merger rate.

\end{abstract}

\maketitle


\section{\label{sec:Intro} Introduction}

With the opening  of the field of gravitational-wave (GW) astronomy and cosmology~\cite{Abbott:2016blz,TheLIGOScientific:2017qsa,LIGOScientific:2020ibl}, we are beginning to use GWs as probes of the  Universe. For some cosmological observables, GWs are complementary  to electromagnetic observations. This is true in particular for the  Hubble parameter $H_0$, for which GWs can provide a measurement with different systematics, that can help to eventually resolve the tension between   
early-Universe~\cite{Aghanim:2018eyx,Abbott:2018xao}
and late-Universe~\cite{Riess:2019cxk,Wong:2019kwg,Riess:2020fzl} measurements of $H_0$. However, there is also cosmological information that can be obtained uniquely through GWs. In general, 
on cosmological scales it is convenient to perform a separation between a
homogeneous background described by a Friedmann-Robertson-Walker (FRW) metric, and  scalar, vector and tensor perturbations over it. The possibility of deviations from GR in the sector of tensor perturbations  can only be investigated using GWs, so this is a genuinely new window that we are beginning to open. In this sector, the most immediate possibility is a deviation of the speed of GWs from the speed of light, but the multi-messenger observation of the binary neutron star (BNS) coalescence 
 GW170817~\cite{TheLIGOScientific:2017qsa,Goldstein:2017mmi,Savchenko:2017ffs,Monitor:2017mdv} has put an extremely strong limit  on it, at the level  $|c_{\rm gw}-c|/c< {\cal O}(10^{-15})$. However, there is much more than the speed of GWs  to be investigated in the  tensor sector. In particular, in the last few years it has been appreciated that
all  modified gravity models that pass the constraint on the speed of GWs still predict a different evolution of the amplitude of GWs in their propagation over cosmological distances.
This is determined by the equation for tensor perturbations over FRW that, in GR, takes the form
\be\label{prophGR}
\tilde{h}''_A  +2 {\cal H} \tilde{h}'_A+c^2k^2\tilde{h}_A=0\, ,
\ee
where $h_{+,\times}$ is the GW amplitude for the two polarizations, the prime denotes the derivative with respect to conformal time $\eta$,   ${\cal H}=a'/a$ and $a(\eta)$ is the FRW scale factor.
In modified gravity, even in  theories that do not change the coefficient of the $c^2k^2$ term (that would result in a different speed of GWs), still the 
 ``friction term" proportional to $\tilde{h}'_A$ is different, and  \eq{prophGR} becomes~\cite{Saltas:2014dha,Lombriser:2015sxa,Nishizawa:2017nef,Arai:2017hxj,Belgacem:2017ihm,Amendola:2017ovw,Belgacem:2018lbp,Belgacem:2019pkk}, 
\be\label{prophmodgrav}
\tilde{h}''_A  +2 {\cal H}[1-\delta(\eta)] \tilde{h}'_A+c^2k^2\tilde{h}_A=0\, ,
\ee
for some function of time $\delta(\eta)$ that encodes the difference from GR. One can then show that, while in GR  the signal from a coalescing binary allows the extraction of  the luminosity distance of the source, in  the context of modified gravity it rather   provides a determination of a different
quantity, $\dgw(z)$, called the ``GW luminosity distance"~\cite{Belgacem:2017ihm}. This is related to the standard luminosity distance [that, in this context, we call the `electromagnetic luminosity distance' and denote by $\dem(z)$] by~\cite{Belgacem:2017ihm,Belgacem:2018lbp}
\be\label{dLgwdLem}
\dgw(z)=\dem(z)\exp\left\{-\int_0^z \,\frac{dz'}{1+z'}\,\delta(z')\right\}\, ,
\ee
where $\delta(z)\equiv \delta[\eta(z)]$.
A useful parametrization of this effect, which catches  the full redshift dependence in terms of just two parameters $(\Xi_0,n)$,  is obtained writing~\cite{Belgacem:2018lbp},
\be\label{eq:fit}
\frac{d_L^{\,\rm gw}(z)}{d_L^{\,\rm em}(z)}\equiv \Xi(z)=\Xi_0 +\frac{1-\Xi_0}{(1+z)^n} \, .
\ee
As shown in ref.~\cite{Belgacem:2019pkk}, \eq{eq:fit} fits remarkably well the explicit results from  typical modified gravity models. GR is recovered 
when $\Xi_0=1$ (for all $n$). The study of explicit modified gravity models shows that $\Xi_0$ can be significantly different from $1$. In particular, in the  RT non-local gravity model  \cite{Maggiore:2013mea} (see \cite{Belgacem:2020pdz} for review),  it can be as large  as $1.80$
\cite{Belgacem:2019lwx,Belgacem:2020pdz}, corresponding to a $80\%$ deviation from  
GR.\footnote{More precisely, the RT model has a free parameter $\Delta N$, related to the choice of initial conditions (defined by starting the evolution during a phase of primordial inflation, $\Delta N$ e-folds before the end of inflation) and, for large $\Delta N$, the prediction for $\Xi_0$  of the RT model saturates to the value $\Xi_0\simeq 1.80$ (and $n\simeq 1.91$). In contrast, if the model is started with initial conditions of order one during radiation dominance, one rather finds $\Xi_0\simeq 0.93$, a $7\%$ deviation from GR.\label{note:inflation}}
 This is quite remarkable, since the deviations from GR and from $\Lambda$CDM for the background evolution and  scalar perturbations are observationally bounded to be much smaller, of the order of at most a few percent. Thus, the newly opened window of tensor perturbations might bring significant surprises.

In GR, where $\dgw(z)=\dem(z)\equiv d_L(z)$, the GW observation gives $d_L(z)$ and, for a source with electromagnetic counterpart, we also have the redshift $z$ of the source. We can then use this to extract $H_0$, as already successfully done in for 
GW170817~\cite{Abbott:2017xzu}. The error from this single detection is still too large to discriminate between the value of $H_0$ obtained from  early and late-Universe probes, and one can estimate  that ${\cal O}(50-100)$ standard sirens with counterpart  are needed to reach the accuracy required  to arbitrate this discrepancy~\cite{Chen:2017rfc,Feeney:2018mkj}.
Binary black holes (BBHs), in general, are not expected to have an electromagnetic counterpart, and also for BNSs only a small fraction of relatively close sources  is expected to have an observed counterpart. 
However, several statistical methods have  been investigated to extract cosmological information from an ensemble of detections. In this context, compact  binary coalescences  without electromagnetic counterpart are often called `dark sirens'. The first and most widely studied possibility is to correlate  
the GW signals with galaxy catalogs~\cite{Schutz:1986gp,DelPozzo:2011yh,Gray:2019ksv,Nair:2018ign,Mukherjee:2019wcg,Yu:2020vyy,Vijaykumar:2020pzn,Mukherjee:2020hyn,Bera:2020jhx,Mukherjee:2020mha}, and has been applied to extract $H_0$ from the recent LIGO/Virgo detections in a number of papers~\cite{Soares-Santos:2019irc,Palmese:2020aof,Abbott:2019yzh,Finke:2021aom, LIGOScientific:2021aug, Palmese:2021mjm}.  Another option, proposed more recently, is to use multiply lensed GW events to identify, or restrict, the host galaxy~\cite{Hannuksela:2020xor,Finke:2021znb}.
It is, however, of great interest to explore possibilities that do not make use of any electromagnetic information, not even in the form of a galaxy catalog. This is particularly relevant for events at high redshift, where the completeness of catalogs and the possibility of detecting a counterpart can be a significant limitation. Since, as we will see, the effect of $\Xi_0$ cumulates with the propagation, this is particularly important for testing modified gravity.

The basic issue with GW observations is that the redshift $z$ of the source enters  in the waveform of a coalescing binary only in the combinations \be\label{mdet_vs_ms}
m^{(\rm det)}_{i}\equiv m_{i} (1+z)\, ,
\ee 
where $i=1,2$ labels the two compact objects in the binary,  $m_{i}$ are the intrinsic (`source-frame') masses, and   $m^{(\rm det)}_{i}$ (the `detector-frame' masses) are  the observed quantities. To determine $z$ from the observed $m^{(\rm det)}_{i}$ one therefore needs further information, involving explicitly a mass scale, that allows us to reconstruct $m_{i}$. For BNSs one possibility is to make use of the fact that, at 5PN order, the waveform depends also on the tidal deformability of the component stars. The latter, given an equation of state, is determined by the intrinsic NS mass. This allows us in principle to break the degeneracy between  $m_{i}$ and $z$~\cite{Messenger:2011gi}. Another option is that an intrinsic mass scale enters through the prior of a Bayesian analysis, rather than through the waveform. For instance, for BNSs, a possibility is to exploit the
narrowness of the neutron star mass distribution~\cite{Taylor:2011fs,Taylor:2012db,Finke:2021eio}. 

For BBHs, the idea put forward in \cite{Farr:2019twy} is to make use of the fact that the mass function of BBHs, whose component BHs originated from stellar collapse, contains explicit mass scales. These reflect the presence of
a mass gap,  approximately in the region $\sim 40-120 \, M_{\odot}$ where, because of the pulsational pair instability supernova (PISN) process, stellar collapse does not produce a BH remnant, but rather an explosion that disperses the stellar material in the interstellar medium.  The presence of the mass scales corresponding to the lower and upper edge of the gap breaks the degeneracy between source-frame mass and redshift, and allows us to perform  a  joint inference on the  cosmological parameter and on the parameters describing the astrophysical population~\cite{Farr:2019twy,Ezquiaga:2020tns,You:2020wju,Mastrogiovanni:2021wsd,Ezquiaga:2021ayr, LIGOScientific:2021aug}.
This method, that will be reviewed in more detail in Sect.~\ref{sec:method},  will be the focus of the present paper. 

All these methods have been first proposed to measure $H_0$ in the context of GR. However, they can be extended to the context of modified gravity. In this case, the GW observations measure $\dgw(z)$ and, in the parametrization (\ref{eq:fit}), the parameter space is therefore enlarged to include $(\Xi_0,n)$. The most general approach is to  perform a joint inference over all cosmological and astrophysical parameters; a simpler approach, for what concerns the cosmological parameters, is to assume a value (or a narrow prior) for $H_0$, such as that given by {\it Planck\,}~\cite{Aghanim:2018eyx}, and focus on the accuracy that can be obtained on $\Xi_0$.\footnote{The parameter $n$ is less important since it only determines the precise shape of the function that interpolates between $\dgw(z)/\dem(z)=1$ at $z=0$ and $\dgw(z)/\dem(z)\simeq \Xi_0$ at large $z$. One could then fix it to typical values suggested by explicit model, such as the value $n\simeq 1.91$  obtained in the RT nonlocal model. Alternatively, one can marginalize also over $n$, which results in limits on $\Xi_0$ larger by about a factor of 2~\cite{Mukherjee:2020mha}. In this paper we will include also $n$ among the parameters determined by the Bayesian inference.} Several results have already been obtained with current data.
Using GW170817 as a standard siren with counterpart, 
a first limit was obtained in~\cite{Belgacem:2018lbp}, using  the electromagnetic luminosity distance of the galaxy hosting the counterpart,  obtained   from surface brightness fluctuations (see also \cite{Lagos:2019kds}, and the discussion in App.~\ref{sec:comparison}, for another analysis using GW170817). Because of the very small redshift 
of GW170817, this is really a limit on $\delta(z=0)$, independently of the parametrization used, and the result from~\cite{Belgacem:2018lbp} is
\be\label{delta0}
\delta(0)=-7.8^{+9.7}_{-18.4}\, .
\ee
Setting for illustration $n=1.91$ (which is the value predicted by the RT model in the same limit in which $\Xi_0=1.80$, see footnote~\ref{note:inflation}) and using $\delta(0)=n(1-\Xi_0)$, this can be translated into 
$\Xi_0=5.1^{+9.1}_{-5.1}$, and therefore into an upper bound
$\Xi_0\,\lsim\, 14$.
This is a very broad limit since, as $z\ra 0$, the effect  of modified GW propagation disappears, see  \eq{dLgwdLem}, and GW170817 has a very small redshift,
$z\simeq 0.01$. 

Recently, in \cite{Finke:2021aom} a much more stringent limit has been obtained,\footnote{This value updates the value quoted in the v1 version of   \cite{Finke:2021aom}, and is obtained improving the analysis with the use of full inspiral-merger-ringdown waveforms, and other technical improvements such as a higher threshold on the local completeness of the catalog used to select GW events.}  
\be\label{Xi0limit1}
\Xi_0=2.1^{+3.2}_{-1.2} \, ,
\ee
($68\%$ c.l.), using  BBH dark sirens from the O1, O2 and O3a LIGO/Virgo run, and performing a correlation  with the GLADE~\cite{Dalya:2018cnd} galaxy catalog. 
This result made however use of a fixed population model, and including the uncertainty on the population would result in a weaker constraint, see e.g. the discussion on page 51 of \cite{Finke:2021aom}.
An even more stringent result is obtained if one accepts the  tentative  identification of the flare  ZTF19abanrhr as the electromagnetic counterpart of the BBH coalescence GW190521. Then, the analysis in \cite{Finke:2021aom} gives  
\be\label{Xi0limit2}
\Xi_0=1.8^{+0.9}_{-0.6}\, ,
\ee 
($68\%$ c.l.). This is consistent with similar results obtained in \cite{Mastrogiovanni:2020mvm} with the same parametrization of modified GW propagation.
Modified GW propagation has also been recently constrained  in~\cite{Ezquiaga:2021ayr}, using  the BBH mass distribution,   following the strategy proposed in \cite{Farr:2019twy} and the GWTC-2 catalog. This will also be the strategy followed in this paper, and in app.~\ref{sec:comparison} we will  compare our results with those in  \cite{Ezquiaga:2021ayr}. 

Several forecasts have also been presented for the accuracy that could  be obtained on $\Xi_0$  in the near future and with third-generation detectors such as  the  Einstein Telescope (ET)~\cite{Punturo:2010zz,Maggiore:2019uih} and Cosmic Explorer~\cite{Reitze:2019iox}, or for  the LISA space interferometer~\cite{Audley:2017drz}. 
In particular, refs.~\cite{Belgacem:2018lbp,Belgacem:2019tbw} provide estimates for the accuracy that could be obtained at ET, using BNSs "with gamma-ray bursts (GRB) as counterparts (see 
\cite{Belgacem:2019zzu,Mastrogiovanni:2020gua} for further studies of BNSs with counterparts), while the perspective for detecting modified GW propagation at 3G detectors from  the reconstruction of the BNSs mass function is discussed in~\cite{Finke:2021eio}; ref.~\cite{Jiang:2021mpd} studies the accuracy that could be obtained on $\Xi_0$ using the tidal deformation of  neutron stars;  forecasts of the accuracy that could be obtained in the future  from BBHs
by performing  cross-correlations with galaxy surveys are presented in~\cite{Mukherjee:2020mha,Canas-Herrera:2021qxs}, while
the possibility of measuring $\Xi_0$ at ground-based detectors with strongly lensed GW events has been discussed in~\cite{Finke:2021znb}.
For LISA, forecasts have been presented 
in \cite{Belgacem:2019pkk}, using  the coalescence of supermassive  black holes (SMBH), and 
in \cite{Baker:2020apq}, correlating between LISA SMBH events with the prediction  for large scale structures of a specific modified gravity model (of the Horndeski class).

In this paper we further elaborate on the possibility of extracting $\Xi_0$ using a joint cosmology-population  analysis  that exploits the mass scales in the BBH mass function~\cite{Farr:2019twy,Ezquiaga:2021ayr}. The goal is twofold: first, to obtain constraints on $\Xi_0$  from the latest GW detections. Second, to analyze in more detail the interplay between the cosmological and population analyses and its potential impact 
on near-future observations, showing in particular that the effect of $\Xi_0$ on population analyses is large.
The plan of the paper is as follows. In sect.~\ref{sec:method} we  recall the basic principles that allow us to extract $\Xi_0$ from a joint hierarchical Bayesian inference of astrophysical and cosmological parameters,  and we introduce the BBH population model that we will use. In sect.~\ref{sec:LVC} we present the results obtained from the GWTC-3 catalog of GW detections~\cite{LIGOScientific:2021djp}, while in sect.~\ref{sec:forecasts} we present forecasts based on 5 years of observations with the advanced LIGO/Virgo detectors, and we study the bias that would be introduced on the population parameters if the data are analysed within GR when the correct theory of Nature has $\Xi_0$ significantly, but plausibly different from the GR value $\Xi_0=1$.
In sect.~\ref{sec:Conclusion} we present our conclusions. Some more technical material is discussed in the appendices.

\section{\label{sec:method}Methods}

As we have mentioned, in the context of modified gravity the luminosity distance measured by GW observations of a coalescing binary is actually the GW luminosity distance $\dgw$.
For a given observed value of $\dgw$, the parameters
$(\Xi_0,n)$ that characterize modified GW propagation affect the 
inferred redshift of the source. In particular, if $\Xi(z)>1$ at all redshifts, $\dgw(z)/\dem(z)>1$, so the electromagnetic luminosity distance is smaller than the observed $\dgw$,  and then also the actual redshift of the source is smaller, compared to the one that would be inferred in GR (for the same values of $H_0$ and $\oma$),\footnote{It should be observed that a modified gravity model, such as the RT nonlocal model, that predicts large deviations of $\Xi_0$ from 1, possibly as large as $80\%$, also fits the cosmological observations with  values of $H_0$ and $\oma$ very close to that of $\Lambda$CDM, with differences at the $0.1\%$ level. Therefore the effect of $\Xi_0$ largely dominates, and this is the situation that we will typically consider.}  while for $\Xi(z)<1$ it is  larger.
This has two main consequences. First, for  a given detector sensitivity, which corresponds to a given range in $\dgw$, the corresponding range in redshift will change. For $\Xi_0>1$, the detector range will be limited to sources at smaller redshift, compared to what would have been possible if $\Xi_0=1$. This can also be understood observing that, for $\delta(z)<0$ [that, according to \eq{dLgwdLem}, implies $\dgw(z)/\dem(z)>1$] the friction term in \eq{prophmodgrav} is larger than in GR, and therefore the GW amplitude is more strongly damped during its propagation, so that only signals from closer sources are detectable; conversely, the reach in $z$ increases with respect to GR if $\Xi_0<1$.
Another important effect is on the reconstruction of the source-frame masses $m_{i}$ from the observed detector-frame masses $m^{(\rm det)}_{i}$.  We have seen that, if the correct theory of Nature is a modified gravity model with $\Xi_0>1$, the true redshift of the source is smaller than the one that would be inferred using GR; then, from \eq{mdet_vs_ms}, we see that, for a given observed $m^{(\rm det)}_{i}$, the actual source-frame masses $m_{i}$ are larger than that inferred using GR. Conversely, for $\Xi_0<1$ they are smaller. 
Information about features in the source-frame mass distribution and merger rate can thus be used to constrain  $\Xi_0$, as a variation in the value of the latter affects the reconstruction of such scales.
Although we focused here on the effect of $\Xi_0$, analogous arguments hold for the effect of $H_0$, for which this method has in fact originally been proposed. The only difference is that, since the luminosity distance is proportional to $1/H_0$, the effect of varying $H_0$ on the reconstruction of mass scales and rates will be inverted with respect to $\Xi_0$ - increasing the value of $H_0$ leads to smaller reconstructed source frame masses and higher rates, and vice versa.

Of course, population and cosmological parameters have to be inferred jointly. This is done with a hierarchical Bayesian approach, which we briefly summarize here (see~\cite{Loredo:2004nn, Adams:2012qw, Mandel:2018mve, Thrane:2018qnx, Vitale:2020aaz}). We denote by $N_{\rm obs}$ the number of events observed, and by $\theta_i$,  with $i=\{1, ..., N_{\rm obs}\}$, the parameters of the $i$-th individual event, with GW data $\mathcal{D}_i$. The data from all observations are denoted by $\mathcal{D}$. Among the parameters, those relevant for this analysis are 
\be\label{thetai}
\theta_i = \{ m^{(\rm det)}_{1,i}, \, m^{(\rm det)}_{2,i}, d^{\rm gw}_{L,i}\} \, ,
\ee 
where $m^{(\rm det)}_{1,i}, \, m^{(\rm det)}_{2,i}$ are the \emph{detector} frame masses of the two component BHs and $d^{\rm gw}_{L,i}$ the GW luminosity distance of the $i$-th event. 
These are the quantities which are directly measured by the GW observation (in contrast to source-frame mass, electromagnetic luminosity distance, or redshift, that have to be inferred using a cosmological model).

The hyperparameters of the model (i.e., the parameters on which we want to draw an inference) are the BBH population parameters $\Lambda_{\rm BBH}$ and the cosmological parameters $\Lambda_{\rm cosmo}$.  We collectively denote them by $\Lambda=\{\Lambda_{\rm BBH}, \Lambda_{\rm cosmo} \}$. 
For the cosmological parameters, we take
\be\label{Lambdacosmo}
\Lambda_{\rm cosmo} = \{H_0, \oma, \Xi_0, n\}\, .
\ee
The BBH population parameters will be specified below. The likelihood is that of an inhomogeneous Poisson process that includes measurement uncertainty and selection biases~\cite{Loredo:2004nn, Mandel:2018mve}, 
\begin{eqnarray}\label{likelihood}
  \mathcal{L}(\mathcal{D} | \Lambda) \propto  \text{e}^{-N_{\rm exp}(\Lambda)}  
  \prod_{i=1}^{N_{\rm obs}} \int d\theta_i \, \mathcal{L}(\mathcal{D}_i| \theta_i)\; 
  \frac{d {N}}{d \theta_i}(\Lambda)\, .
 \end{eqnarray}
 The three building blocks of this expression are  $dN/d\theta_i$, $\mathcal{L}(\mathcal{D}_i| \theta_i)$, and $N_{\rm exp}$. We discuss them in turn.
 
The term $dN/d\theta_i$ describes  the population and is given by 
\bees
\frac{d {N}}{d \theta}(\Lambda) &=&  \frac{d {N}(\Lambda)}{d m^{(\rm det)}_1 d m^{(\rm det)}_2 d (\dgw) }\\
&=& \frac{1}{(1+z)^2 \,  \frac{d (\dgw)}{d z}(\Lambda_{\rm cosmo}) } \, 
\frac{d {N}}{d m_1 d m_2 d z }(\Lambda_{\rm BBH})\, .\nn
\ees
The factor $(1+z)^2$ is the Jacobian between source-frame masses $m_i$  and the detector-frame masses
$m^{(\rm det)}_i$, while, using \eq{eq:fit} and the standard expression of the electromagnetic luminosity distance,
\be\label{dLemLCDM}
d_L(z)=(1+z)\dcom(z)=
c (1+z) \, \int_0^z\, 
\frac{d\tilde{z}}{H(\tilde{z})}\, ,
\ee
the transformation from  luminosity distance and redshift is given by
\begin{equation}\label{ddgwdz}
\frac{d (\dgw)}{d z} = \Big[ \Xi(z) - \frac{n \ (1-\Xi_0) }{(1+z)^{n}}  \Big] d_{\rm com}(z) 
+ \frac{(1+z) \ \Xi(z)}{H(z)} \, .
\end{equation}
In the above equation, $d_{\rm com}(z)$ is the comoving distance and $\Xi(z)$ is defined in \eq{eq:fit}. 
The quantity $dN/(dm_1 dm_2 dz)$ describes the population as a differential  with respect to its most natural variables, the redshift and the source-frame masses and, as we will see explicitly below,  it depends on the astrophysical hyperparameters $\Lambda_{\rm BBH}$. However, the cosmological parameter  enter in $dN/[d m^{(\rm det)}_1 d m^{(\rm det)}_2 d (\dgw) ]$ because of the transformation from $\dgw$ to $z$, that involves both the  parameters $(\Xi_0,n)$, that appear explicitly in \eq{ddgwdz}, and the standard cosmological parameters such as $H_0$ and $\oma$ that enters in 
\eq{ddgwdz} through $\dcom(z)$. Finally, a dependence on $\Lambda_{\rm cosmo}$ is always present in the redshift $z$, since this is obtained from the observed GW luminosity distance for a given set of cosmological parameters through the inversion of the equation $d_L^{\rm gw,  obs} =  \dgw(z; \Lambda_{\rm cosmo})$  with the right hand side given by \eqs{eq:fit}{dLemLCDM}. 

The differential mass-redshift distribution of BBHs with respect to the source parameters is written as~\cite{LIGOScientific:2018jsj}
\bees
\frac{dN}{dm_1dm_2dz}(\Lambda_{\rm BBH})&=& \[ \frac{dV_c}{dz}(z)\]
\, \frac{T_{\rm obs}}{1+z}\nn\\
&&\times \,R(z|\Lambda_z)\, p(m_1,m_2|\Lambda_m)\, ,
\ees
where  $T_{\rm obs}$ is the total observation time, $dV_c/dz$ is the differential comoving volume per unit redshift, $R(z|\Lambda_z)$ is the merger rate density, and 
$p(m_1,m_2|\Lambda_m)$ is the  distribution of source-frame component masses. We denoted by  $\Lambda_m$  the hyperparameters that describe the mass distribution, and by $\Lambda_z$  the hyperparameters that enter in the merger rate density; assuming that the mass distribution does not evolve with redshift, these are disjoint sets,  so $\Lambda_{\rm BBH}=\{\Lambda_m,\Lambda_z\}$. 
The assumption that the mass distribution is not evolving with redshift is consistent with the GWTC-2 events for the mass model assumed in this work~\cite{Fishbach:2021yvy}, and the expected dependence is mild, in particular at the redshifts covered by GWTC-3~\cite{Mapelli:2019bnp}. Finally, we  have neglected the dependence on the spins.

For $R(z|\Lambda_z)$, we use the
Madau-Dickinson parametrization~\cite{Madau:2014bja,Madau:2016jbv}
\be\label{Madau}
R(z|\Lambda_z) = R_0 \,  \left[1 + (1+z_p)^{-\gamma-k} \right]\,  \frac{(1+z)^{\gamma}}{1 + \left (  \frac{1+z}{1+z_p}   \right )^{\gamma + k}}\, .
\ee
Therefore
\be\label{Lambdaz}
\Lambda_z=\{R_0, \gamma, k, z_p\}\, .
\ee 
The redshift  $z_p$ is the peak of the star formation rate (which is in the range $z_p\sim 2-3$) and  $\gamma$ and $k$ are constants. This functional form  interpolates between a power-like behavior $R(z)\propto (1+z)^{\gamma}$ for redshifts well below $z_p$, and $R(z)\propto (1+z)^{-k}$ at $z\gg z_p$. 

For $p(m_1,m_2|\Lambda_m)$ we use the broken power-law  model described in App.~B3 of
\cite{LIGOScientific:2020kqk}, which interpolates between two power laws with different slopes, matched at a ``breaking point'' $m_\text{break}$. The precise definition is given in App.~\ref{sec:massfunction}, and has 7 parameters,
\be\label{Lambdam}
\Lambda_m=\{\alpha_1, \alpha_2, \beta_q, \delta_m, m_{\rm min}, m_{\rm max}, b \} \, ,
\ee
with the scale $m_\text{break}$ being defined as 
\be
m_\text{break} = \mmin +b(\mmax-\mmin) \, .
\ee
In conclusion, together with \eq{Lambdaz}, we get
\be\label{LambdaBBH}
\Lambda_{\rm BBH}=\{R_0, \gamma, k, z_p,\alpha_1, \alpha_2, \beta_q, \delta_m, m_{\rm min}, m_{\rm max}, b \}\, .
\ee
The broken power-law model  for the mass distribution is the simplest one to account for the presence of a scale in the mass distribution without introducing a sharp cutoff at the lower edge ($\sim 45\msun$) of the PISN mass gap, as for instance in the ``truncated power law'' mass model described in app.~A of \cite{Farr:2019twy}. This is needed since a few detections imply a non-zero merger rate beyond this mass limit.  The mass scale $m_\text{break}$ can be thought of as modeling the onset of the PISN instability, and the second power law can be though either as a gradual smoothing of the lower edge of the PISN instability, or due to a subpopulation of BBHs within the PISN mass gap~\cite{LIGOScientific:2020kqk}.
In any case, the presence of a mass scale such as  $m_\text{break}$ is what we need to break the degeneracy between source-frame masses and redshifts, as discussed in sect.~\ref{sec:Intro}. 
As shown in \cite{LIGOScientific:2020kqk}, the broken power-law model fits well the data from the GWTC-2 catalog, while in the GWTC-3 catalog the presence of more features has been investigated  \cite{LIGOScientific:2021psn}, but at the current level a model accounting for one feature is enough for cosmological analysis \cite{LIGOScientific:2021aug}.

The second building block of  \eq{likelihood}, $\mathcal{L}(\mathcal{D}_i| \theta_i)$, is the GW likelihood for the $i$-th event. In practice, in a hierarchical inference one has rather access to a discrete set of samples from the posterior distribution for $\theta_i$ for each event, which are related to the likelihood by Bayes' theorem and thus include the prior $\pi(\theta_i)$ used in the parameter estimation.
The marginalization over the source parameters in \eq{likelihood} can be achieved by importance sampling, computing the average of the population function $dN/d\theta$ over the posterior samples weighted by the prior.
In particular, the LIGO/Virgo Collaboration (LVC) uses flat priors in detector-frame masses and a prior on luminosity distance $\pi(d_L)\propto d_L^2$.\footnote{Here $d_L$ is just the value of each MC sample of the parameter estimation. In the context of modified GW propagation we interpret it as $\dgw$, but this is not relevant here.} This leads to the following posterior distribution for $\Lambda$:
\begin{eqnarray}\label{posterior0}
  p(\Lambda | \mathcal{D} ) \propto \pi(\Lambda )\,  \text{e}^{-N_{\rm exp}(\Lambda)}  \prod_{i=1}^{N_{\rm obs}} \left\langle\frac{1}{\pi(\theta_i)}  \frac{d {N}}{d \theta_i}(\Lambda) \right\rangle_{\rm samples} 
\end{eqnarray}
where $\pi(\Lambda )$ is the prior on the hyperparameters.

Finally, $N_{\rm exp}(\Lambda)$ in \eq{likelihood} and \eq{posterior0} represents the \emph{expected} number of detections and corrects for selection bias. We compute it with the weighted MC integration method introduced in \cite{Tiwari:2017ndi} and  described in detail in App.~\ref{sec:SelBias}.  
We also include a small correction to \eq{posterior0} that accounts for uncertainty in the MC computation of $N_{\rm exp}(\Lambda)$ \cite{Farr_2019}  as described in App.~\ref{sec:SelBias}, which leads to the posterior in \eq{posteriorFull}.

\begin{table*}[th]
 \begin{tabular}{@{}ccccccccccccccccccccc@{}}
 \toprule
  \textbf{Parameter}  & $n$  & $R_0$ & $\gamma$ & $k$ & $z_p$   & $\alpha_1$ & $\alpha_2$ & $\beta_q$ &  $\delta_m$ & $m_{\rm min}$& $m_{\rm max}$ & $b$  \\ 
  \hline
   \textbf{Type of prior} &  flat   &    flat & flat & flat & flat   & flat & flat & flat &  flat & flat     &     flat & flat  \\ 
\hline
  \textbf{Range}  &$(1,5)$ & $(0, 100)$  & $(0, 12)$ & $(0, 6)$ & $(0, 4)$ & $(1.5, 12)$ & $(1.5, 12)$ & $(-4, 12 )$ & $(0, 10)$ & $(2, 50)$ & $( 50, 200)$ & $(0, 1)$\\
  \botrule
 \end{tabular}

 \caption{Summary of prior choices. $H_0$ and $\oma$ are fixed to Planck 2018 values in the main analysis, while results for $H_0$ and $\oma$ with flat priors and at fixed $(\Xi_0, n)$ are given in App.~\ref{sec:H0}. The choice of the prior range for $m_{\rm max}$ is discussed in the text, and comparison with a narrower range $m_{\rm max} \in [50, 100]$ is discussed in App.~\ref{sec:mmax}. For $\Xi_0$, as discussed in the text, we consider  a flat prior and a prior flat in $\log \Xi_0$, both between $[0.1, 10]$.
  $R_0$ is in units of ${\rm Gpc}^{-3}\,{\rm yr}^{-1}$ and $\mmin$, $\mmax$ are in units of $\msun$. }
\label{tab:priors}
\end{table*}

\section{\label{sec:LVC}Results from GWTC-3}

In this section we present the results obtained by analyzing the GWTC-3 catalog~\cite{LIGOScientific:2021djp}. 
\subsection{Data and prior choices}\label{sect:data}

For the main results, we restrict ourselves to a ``pure'' sample of events with network matched-filter signal-to-noise ratio $\text{SNR}>12$. We also exclude the events GW170817, GW190425, GW190426$\_$152155, GW190814, GW200115$\_$042309, GW191219$\_$163120, GW200210$\_$092254, GW190917$\_$114630 because for all of them at least one of the component masses is compatible with that of a neutron star (NS). This results in a sample of 35 events. A comparison with other choices of threshold in SNR is performed in App.~\ref{sec:SNRcuts}.

The event GW190521~\cite{Abbott:2020tfl,Abbott:2020mjq} has a special role. It is the most massive event with a confident detection in the GWTC-3 catalog, with masses
$m_1=91.4^{+29.3}_{-17.5}\, \msun$ and $m_2=66.8^{+20.7}_{-20.7}\, \msun$ that, particularly for the primary mass $m_1$, fall inside the PISN mass gap with high statistical confidence.
The analysis in ref.~\cite{LIGOScientific:2020ibl} cannot determine if it belongs to a distinct subpopulation or to the high-mass tail of the BBH population considered in the analysis, and the nature of this event remains debated \cite{Gayathri:2020coq,Nitz:2020mga,Romero-Shaw:2020thy,Estelles:2021jnz,Hu:2021lbt}. In the framework of the broken power-law model that we are also using, it can be interpreted as a normal member of the population (while it is clearly an outlier of the `truncated' mass distribution used to fit the GWTC-1 catalog).  The underlying hypothesis, for this interpretation, is that  the lower edge of the PISN mass gap is not sharp, and events with $m_1\, \gsim\, 45\msun$ belong to the high-mass tail of the BBH population. However, as remarked also in ref.~\cite{LIGOScientific:2020ibl}, this explanation might pose challenges to our understanding of stellar evolution, since the lower edge of the PISN mass gap is believed to be quite sharp~\cite{Woosley:2016hmi,Farmer:2019jed}. 
Since the presence of outliers can be particularly relevant for this analysis, as discussed in more detail below, we will perform it both including and excluding GW190521.

We make use of the publicly available posterior samples labeled $\tt{Overall\_posterior}$ for the GWTC-1 events, of those labeled by $\tt{PublicationSamples}$ for GWTC-2, and of those labeled by $\tt{ImrPhenomXPHM}$ for GWTC-3.\footnote{The samples for GWTC-1, GWTC-2 and GWTC-3 are obtained from  \url{https://dcc.ligo.org/LIGO-P1800370/public/}, \url{https://dcc.ligo.org/LIGO-P2000223/public/}, and \url{https://zenodo.org/record/5546663\#.YaT5jC2ZOL5} respectively.} 
Selection effects are computed as described in App.~\ref{sec:SelBias}.

The full hyperparameter space is given by $\Lambda=\{\Lambda_{\rm BBH}, \Lambda_{\rm cosmo} \}$ with $\Lambda_{\rm BBH}$ given in \eq{LambdaBBH} and $\Lambda_{\rm cosmo}$ given in \eq{Lambdacosmo}. We will, however, also consider subsets of this full parameter space, see below.

Our prior choices are summarized in table~\ref{tab:priors}. One important caveat on the choice of the prior is represented by the allowed range for the maximal mass $m_{\rm max}$. As our base configuration, we choose a large prior range $m_{\rm max} \in [50, 200] \msun$. However, if the underlying physical hypothesis is that the BBHs in the sample are originated from stellar collapse and that their distribution features a ``mass gap'', one 
could be tempted to adopt a more restrictive prior, e.g. $m_{\rm max} \in [50, 100] \msun$, which would correspond to the strong physical assumption that no BHs with masses larger than $100 \msun$ exist in the sample. 
In turn, this choice would make the analysis more sensitive to the presence of population outliers at high mass, potentially resulting in a bias. Keeping in mind the discussion on the role of $\Xi_0$ in Sect.~\ref{sec:method}, the reason is easily understood: low values of $\Xi_0$ lead to lower reconstructed source-frame mass, hence the presence of events with posterior support for the primary mass close to (or above) the upper limit of the prior range would strongly favor values of $\Xi_0<1$. This can be in particular the case for GW190521, whose primary mass largely falls in the expected gap and which does have posterior support above $100 \msun$, as discussed above. So, in case a narrow prior on $m_{\rm max}$ is used, it is particularly important to check the robustness of the result with respect to the inclusion of the potential outlier GW190521. We discuss the choice of a narrower prior in App.~\ref{sec:mmax}. In general, we believe it is better to allow wider prior ranges in order not to rely on too strong (and potentially wrong) physical assumptions.
Finally, we will present the results for  a prior flat in $\Xi_0$ and for a prior flat in $\log\Xi_0$, both between $[0.1, 10]$.
We will see below that, with a sufficiently large threshold in signal-to-noise ratio, such as ${\rm SNR} > 12$, the analysis becomes robust against the choice of priors and the inclusion of GW190521, with the constraint being driven by the mass scale $m_{\rm break}$ whose presence and location remain solid.

We sample \eq{posteriorFull} using the ensemble sampler $\tt{emcee}$~\cite{Foreman-Mackey:2012any} with 30 walkers. Our chains are always longer than 50 times the average autocorrelation time $\hat{\tau}$, with a burn-in phase of $2\hat{\tau}$ being discarded,  and the remaining part further thinned by $\hat{\tau}/2$.

\begin{figure}[t]
\centering
\includegraphics[width=0.5\textwidth]{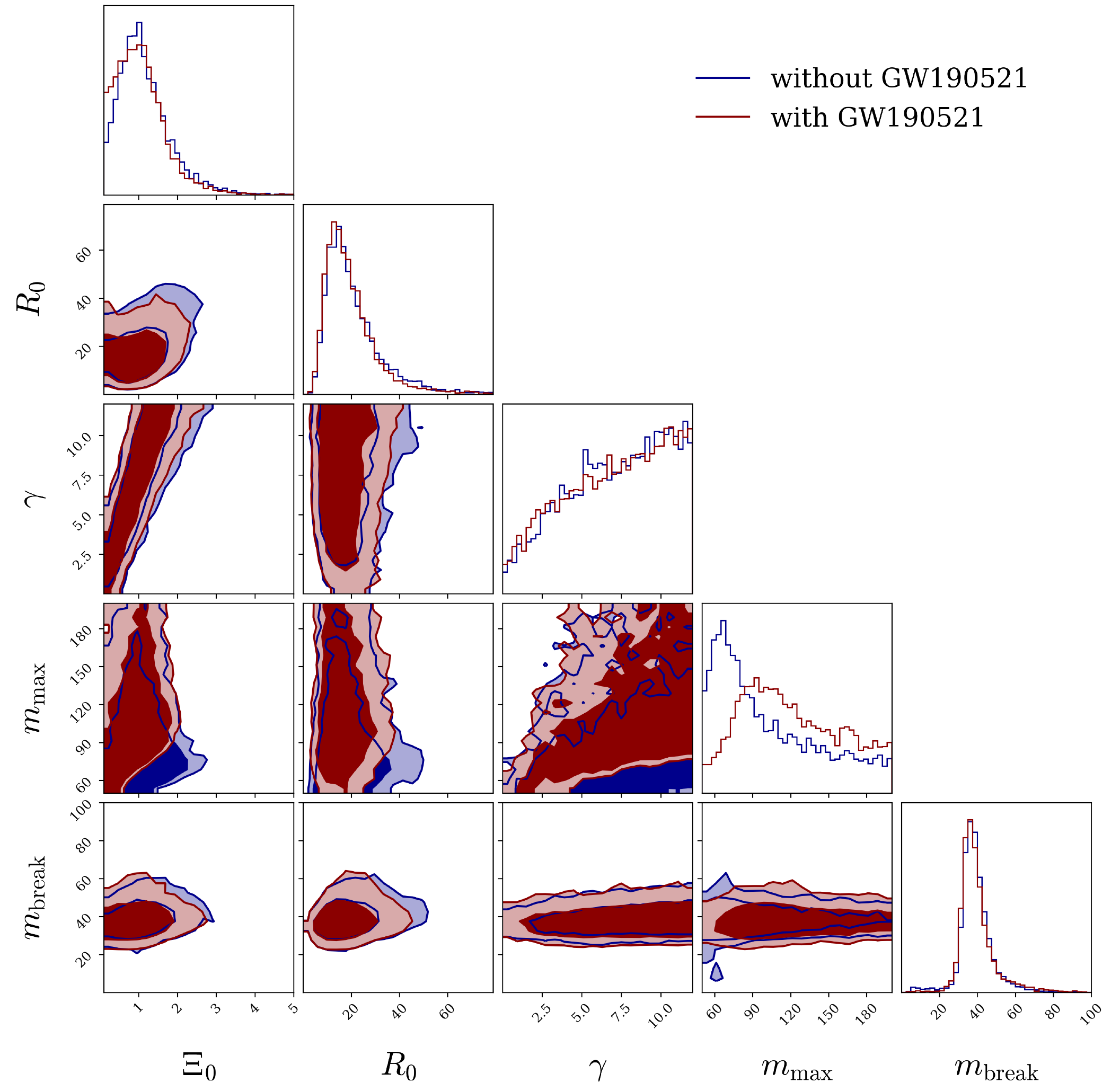}
\caption{Joint constraints from GWTC-3 on the parameters $ \{\Xi_0, R_0, \gamma,m_{\rm max}, m_{\rm break} \}$, with $H_0, \oma$ fixed to the Planck 2018 values and marginalized over all the other parameters. Contours show the $68\%$ and $90\%$ confidence intervals.} 
\label{fig:GWTC2corner}
\end{figure}

\subsection{Joint  population and cosmological inference}\label{sect:fullcosmopop}

  \begingroup
 \squeezetable 
\begin{table*}[th]
 \centering
 \begin{ruledtabular}
 \begin{tabular}{@{}ccccccccccccccc@{}}
  & &  $ \bf \Xi_0$  & $\bf n$ & $\bf R_0$ & $\bf \gamma$  & $\bf k$ & $\bf z_p$ &  $\bf{\alpha_1}$  & $\bf \alpha_2 $ & $\bf \beta_q$ & $\bf {\delta_m}$  & $\bf {m_{\rm min}}$& $\bf {m_{\rm max}}$ & $\bf {m_{\rm break}}$ \\
  \hline
   \hline \\ [-1.5ex]
   
   
\multicolumn{15}{ c }{Flat-in-log prior}\\
   \hline
   \hline \\ [-1.5ex]
  \multirow{2}{*}{\textbf{68\% sym}}&  \textbf{\makecell{With \\ GW190521}} & $1.0^{+0.6}_{-0.5}$ & $2.5^{+1.7}_{-1.1}$ & $17.1^{+11.1}_{-6.6}$ & $7.5^{+3.2}_{-4.1}$ & $3.0^{+2.1}_{-2.1}$ & $2.5^{+1.0}_{-1.1}$ & $2.8^{+0.6}_{-0.6}$  & $7.5^{+2.2}_{-1.7}$ & $0.7^{+1.4}_{-1.2}$& $4.7^{+2.9}_{-3.1}$ & $4.3^{+1.2}_{-1.3}$ & $115.1^{+51.6}_{-33.7}$ & $37.7^{+8.4}_{-4.9}$  \\
   \cline{2-15}
   & \textbf{\makecell{Without \\ GW190521}}  & $1.0^{+0.7}_{-0.5}$ & $2.6^{+1.7}_{-1.3}$& $18.0^{+13.2}_{-7.0}$ & $7.6^{+3.1}_{-3.9}$& $3.0^{+2.0}_{-2.1}$ & $2.4^{+1.1}_{-1.1}$ & $2.7^{+0.6}_{-0.6}$ & $8.1^{+2.5}_{-2.4}$ & $0.7^{+1.4}_{-1.1}$ & $4.6^{+3.0}_{-3.1}$ & $4.2^{+1.3}_{-1.2}$& $89.6^{+62.4}_{-26.6}$& $38.2^{+7.3}_{-5.0}$\\
   \hline
   \hline \\ [-1.5ex]
  \multirow{2}{*}{\textbf{90\% sym}}&  \textbf{\makecell{With \\ GW190521}} & $1.0^{+1.3}_{-0.8}$ & $2.5^{+2.3}_{-1.4}$ & $17.1^{+24.2}_{-9.4}$ & $7.5^{+4.1}_{-5.9}$ & $3.0^{+2.7}_{-2.7}$ & $2.5^{+1.3}_{-1.7}$ & $2.8^{+1.0}_{-0.9}$  & $7.5^{+3.6}_{-2.7}$ & $0.7^{+2.6}_{-1.9}$& $4.7^{+4.3}_{-4.2}$ & $4.3^{+1.7}_{-1.9}$ & $115.1^{+73.9}_{-49.6}$ & $37.7^{+21.7}_{-8.1}$  \\
   \cline{2-15}
   & \textbf{\makecell{Without \\ GW190521}}  & $1.0^{+1.4}_{-0.7}$ & $2.6^{+2.1}_{-1.5}$& $18.0^{+28.5}_{-9.9}$ & $7.6^{+4.0}_{-5.8}$& $3.0^{+2.8}_{-2.7}$ & $2.4^{+1.4}_{-1.6}$ & $2.7^{+1.0}_{-0.9}$ & $8.1^{+3.4}_{-4.2}$ & $0.7^{+2.5}_{-1.8}$ & $4.6^{+4.4}_{-4.2}$ & $4.2^{+1.7}_{-1.9}$& $89.6^{+95.0}_{-34.6}$& $38.2^{+16.1}_{-9.0}$\\
  \hline
   \hline \\ [-1.5ex]
   
   
  \multicolumn{15}{ c }{Flat prior}\\
   \hline
   \hline \\ [-1.5ex]
  \multirow{2}{*}{\textbf{68\% sym}}&  \textbf{\makecell{With \\ GW190521}} & $1.3^{+0.9}_{-0.5}$ & $2.7^{+1.6}_{-1.3}$ & $19.2^{+17.7}_{-7.8}$ & $9.1^{+2.1}_{-3.6}$ & $3.0^{+2.0}_{-2.1}$ & $2.5^{+1.0}_{-1.1}$ & $2.8^{+0.6}_{-0.6}$  & $7.2^{+2.2}_{-1.5}$ & $0.8^{+1.5}_{-1.2}$& $4.8^{+2.8}_{-3.0}$ & $4.3^{+1.2}_{-1.3}$ & $125.0^{+45.7}_{-34.4}$ & $38.9^{+9.5}_{-5.4}$  \\
   \cline{2-15}
   & \textbf{\makecell{Without \\ GW190521}}  & $1.4^{+0.9}_{-0.6}$ & $2.8^{+1.5}_{-1.3}$ & $20.6^{+15.9}_{-8.3}$ & $8.8^{+2.4}_{-3.7}$ & $3.0^{+2.1}_{-2.0}$ & $2.4^{+1.0}_{-1.1}$ & $2.7^{+0.6}_{-0.5}$  & $7.8^{+2.4}_{-2.3}$ & $0.8^{+1.5}_{-1.2}$& $4.9^{+2.7}_{-3.0}$ & $4.3^{+1.2}_{-1.3}$ & $87.8^{+60.0}_{-22.0}$ & $39.5^{+9.5}_{-5.6}$\\
   \hline
   \hline \\ [-1.5ex]
   
  \multirow{2}{*}{\textbf{90\% sym}}&  \textbf{\makecell{With \\ GW190521}} & $1.3^{+1.9}_{-0.8}$ & $2.7^{+2.1}_{-1.6}$ & $19.2^{+32.5}_{-10.9}$ & $9.1^{+2.7}_{-6.1}$ & $3.0^{+2.7}_{-2.7}$ & $2.5^{+1.4}_{-1.6}$ & $2.8^{+1.0}_{-0.9}$  & $7.2^{+3.7}_{-2.5}$ & $0.8^{+2.6}_{-1.9}$& $4.8^{+4.3}_{-4.2}$ & $4.3^{+1.7}_{-1.9}$ & $125.0^{+64.2}_{-28.8}$ & $38.9^{+25.7}_{-8.3}$\\
   \cline{2-15}
   & \textbf{\makecell{Without \\ GW190521}}  & $1.4^{+2.0}_{-0.9}$ & $2.8^{+2.0}_{-1.7}$ & $20.6^{+34.7}_{-11.9}$ & $8.8^{+2.9}_{-5.9}$ & $3.0^{+2.7}_{-2.7}$ & $2.4^{+1.4}_{-1.6}$ & $2.7^{+0.9}_{-0.9}$  & $7.8^{+3.6}_{-4.0}$ & $0.8^{+2.8}_{-1.8}$& $4.9^{+4.1}_{-4.3}$ & $4.3^{+1.7}_{-1.9}$ & $87.8^{+93.7}_{-30.1}$ & $39.5^{+20.5}_{-9.6}$ \\
  
 \end{tabular}
  \end{ruledtabular}
\caption{Median and 68\% (top) - 90\%(bottom) symmetric C.I. for all the parameters used in the full analysis, with the prior choices of Table~\ref{tab:priors} and using the 35 events with ${\rm SNR} > 12$, with and without including GW190521, and for both prior choices on $\Xi_0$. 
$R_0$ has units of $\text{Gpc}^{-3} \text{yr}^{-1}$; $m_{\rm min}$, $m_{\rm max}$ and $m_{\rm break}$ have units of $\msun$.}
\label{tab:constraintssym}
\end{table*}
\endgroup

\begin{figure}[t]
\centering
\includegraphics[width=0.5\textwidth]{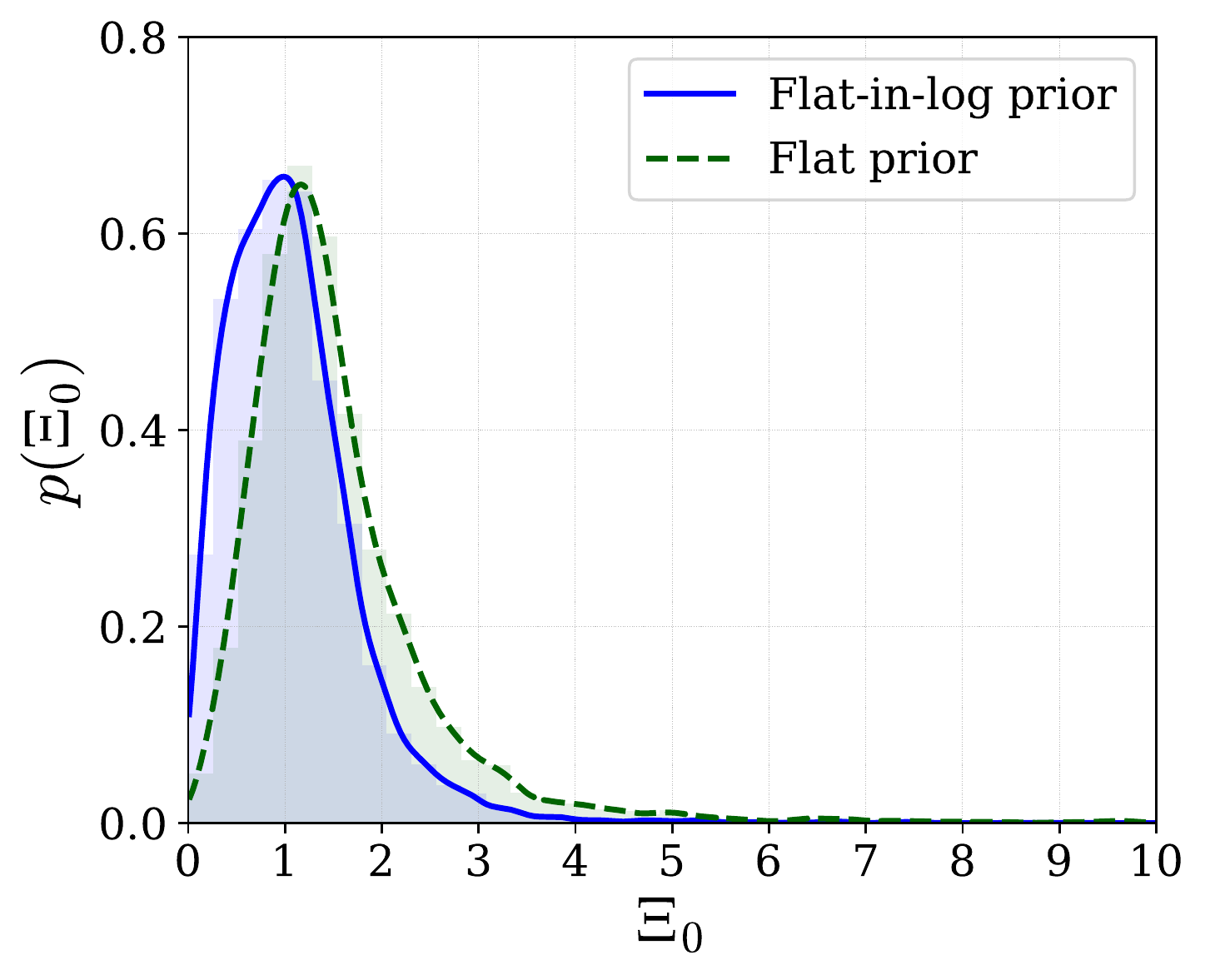}
\caption{Marginal posterior on $\Xi_0$ with a prior uniform in log space (blue, solid line) and a uniform prior (green, dashed line), with GW190521 included in the analysis.}
\label{fig:GWTC3Xi0FlatvsLog}
\end{figure}
 
 \begin{figure}[h]
\centering
\includegraphics[width=0.5\textwidth]{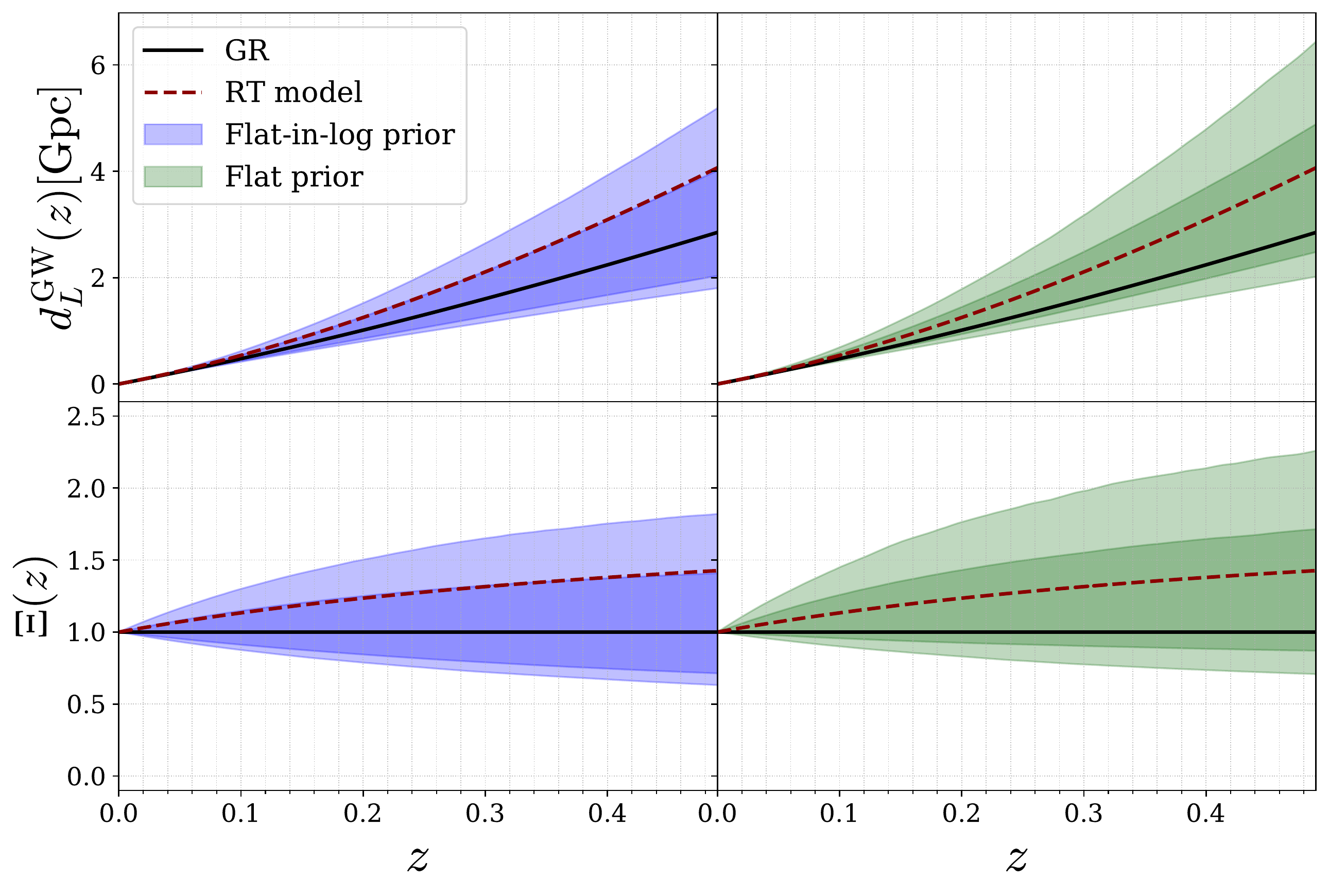}
\caption{Constraint on the GW luminosity distance (upper panel), and on the function $\Xi(z)$ in \eq{eq:fit}, i.e. the ratio between GW and EM luminosity distances (lower panel), obtained from the population analysis of GWTC-3 with $H_0, \oma$ fixed to the Planck 2018 values, and including the event GW190521. The left column (blue) corresponds to the analysis with a prior flat in $\log\Xi_0$, and the right column (green) to the analysis with a prior flat in $\Xi_0$. 
The shaded areas indicate $68\%$ and $90\%$ confidence regions.
We also show  the GR prediction and that of the RT model with $\Xi_0=1.8$ and $n=1.9$ \cite{Belgacem:2020pdz}.
}
\label{fig:GWTC3Xiz}
\end{figure}

\begin{table*}[t]
\centering
 \begin{tabular}{@{}ccccc@{}}
 \toprule
 &  $\bf{\Xi_0^{HDI}} \textbf{(68\%)}$ &  $\bf{\Xi_0^{HDI}} \textbf{(90\%)} $ &  $\bf{\Xi_0^{sym} (68\%)}$  &  $\bf{\Xi_0^{sym} (90\%)}$   \\
  \hline
  \hline \\ [-1.5ex]
\multicolumn{5}{ c }{Flat-in-log prior}\\
\hline
  \textbf{\makecell{With \\ GW190521}} & $1.0^{+0.4}_{-0.8}$ & $1.0^{+0.9}_{-0.9}$ & $1.0^{+0.6}_{-0.5} $ & $1.0^{+1.3}_{-0.8} $ \\
   \hline
   \textbf{\makecell{Without \\ GW190521}}  & $1.0^{+0.5}_{-0.7}$ & $1.0^{+1.2}_{-0.8}$ &  $1.0^{+0.7}_{-0.5}$  &  $1.0^{+1.4}_{-0.7} $\\
   \hline
   \hline \\ [-1.5ex]
    \multicolumn{5}{ c }{Flat prior}\\
   \hline
   \textbf{\makecell{With \\ GW190521}} &  $1.2^{+0.7}_{-0.7}$ & $1.2^{+1.5}_{-1.0}$ & $1.3^{+0.9}_{-0.5}$ & $1.3^{+1.9}_{-0.8}$\\
    \hline
    \textbf{\makecell{Without \\ GW190521}} & $1.2^{+0.8}_{-0.6}$ & $1.2^{+1.6}_{-0.9}$  &$1.4^{+0.9}_{-0.6}$ & $1.4^{+2.0}_{-0.9}$ \\
    \hline
   \hline
  \end{tabular}
 \caption{Marginal constraints on $\Xi_0$ from the main analysis. The first two columns contain the maximum posterior and $68\%$ and $90\%$ HDI. The last two columns contain the median and $68$ and $90\%$ symmetric intervals.}
 \label{tab:constraintsXi0}
 \end{table*}

For the main analysis presented in this section, $\Lambda$ is given by the full set $\Lambda=\{\Lambda_{\rm BBH}, \Lambda_{\rm cosmo} \}$ with $\Lambda_{\rm BBH}$ given in \eq{LambdaBBH} and $\Lambda_{\rm cosmo}$ by $(\Xi_0,n)$, with $H_0$ and $\oma$ fixed to the Planck 2018 mean values $H_0=67.74 \, \text{km} \, \text{s}^{-1} \, \text{Mpc}^{-1}$ and  $\oma=0.311$\footnote{We checked that marginalising on these parameters with Gaussian priors corresponding to the Planck 2018 TT, TE, EE+lowE+lensing+BAO results has a negligible impact on our results.} . For comparison with~\cite{LIGOScientific:2021aug}, in App.~\ref{sec:H0} we report our results for $H_0$ and $\oma$, with $\Xi_0$  fixed at the GR value $\Xi_0=1$.

In Fig.~\ref{fig:GWTC2corner} we show the constraints in the subspace of the full parameter space where there are correlations with physically relevant meaning, i.e. $ \{\Xi_0, R_0, \gamma, m_{\rm max}, m_{\rm break} \}$, using a prior flat in $\log\Xi_0$.
Fig.~\ref{fig:GWTC3Xi0FlatvsLog} shows  the posterior for $\Xi_0$ (with GW190521 included)
both with a flat-in-log prior (blue solid line, the same as the red line in the upper-left panel of Fig.~\ref{fig:GWTC2corner}), and with a flat prior on $\Xi_0$ (green, dashed), with limits $\Xi_0 \in [0.1, 10]$ for both. The joint constraints for the other parameters in the case of a flat prior are very similar to those in Fig.~\ref{fig:GWTC2corner} and we do not show them here. 

Table~\ref{tab:constraintssym} summarizes all the constraints using the median and symmetric confidence intervals (C.I.). Given the high statistical uncertainty, different choices to quantify the credible intervals can result in slightly different values. For parameters where the posterior exhibits a clear peak, a more sound choice is the use of the maximum posterior and  highest-density interval (HDI), so we report also the values obtained in this case for the  constraint on $\Xi_0$, marginalized over all other parameters. Results for $\Xi_0$ with these different choices are reported in Table~\ref{tab:constraintsXi0}. 
Finally, Fig.~\ref{fig:GWTC3Xiz} shows the constraint on the functions $\dgw(z)$ and $\Xi(z)$ obtained including GW190521. As discussed below, the result without GW190521 is basically indistinguishable. %

Comparing the results with and without GW190521 we observe that the presence of GW190521 only impacts $m_{\rm max}$ while the other parameters remain consistent, see Fig.~\ref{fig:GWTC2corner} and Table~\ref{tab:constraintssym}. In particular, we see that the feature in the mass distribution parametrized by $m_{\rm break}$ is constrained to be between $\sim 30 \msun$ and $\sim 45 \msun$, consistently with the expectation from the PISN. More specifically, we get $m_{\rm break}=37.7^{+8.4}_{-4.9}\, \msun$ (median and $68\%$ symmetric C.I., using a flat-log prior on $\Xi_0$ and including GW190521).
The presence of this feature drives the constraint on $\Xi_0$, which, importantly, is not affected by the presence of GW190521. Verifying this explicitly is important in light of the large effect that a potential outlier could have on $\Xi_0$ through the parameter $m_{\rm max}$, as discussed in Sec.~\ref{sect:data}.

The resulting constraints on $\Xi_0$, given in Table~\ref{tab:constraintsXi0} for different  choices (with or without GW190521, and with prior flat in $\Xi_0$ or flat in $\log\Xi_0$),
are basically independent of GW190521 and are the most stringent to date. In particular, they significantly improve on the $\gtrsim 100\%$ constraint obtained in \cite{Finke:2021aom} by correlation with galaxy catalogs, and are even comparable to the bound  $\Xi_0 = 1.8^{+0.9}_{-0.6}$ obtained in \cite{Finke:2021aom} by assuming that the flare ZTF19abanrhr is the electromagnetic counterpart of GW190521, which, however, is not well established \cite{Palmese:2021wcv, Ashton:2020kyr}. The results in Table~\ref{tab:constraintsXi0}  can be also compared to the result found in  ref.~\cite{Ezquiaga:2021ayr}, which basically uses the same methodology (and the GWTC-2 catalog, always including GW190521), but a different parametrization for modified GW propagation. In App.~\ref{sec:cmparam} we discuss the translation between the two parametrizations and,  in App.~\ref{sec:comparison}, we  perform the comparison of the numerical results,  discussing the different assumptions used.

We also observe a positive correlation  in $(\Xi_0,\gamma)$ and in $(\Xi_0,R_0)$: this is due to the fact that increasing $\Xi_0$ pushes events with given $z$ at higher $\dgw$, so less events would be detected, and we  need a higher rate $R(z)$ to reproduce the observed number of events. This can be obtained  increasing $R_0$, or (at the redshifts $z\,\lsim\,  z_p$ of the current events) increasing $\gamma$. The latter is also affected by the evolution of the function $\Xi(z)$ with redshift, encoded in the parameter $n$. Because of this degeneracy, the error on $\gamma$ is very large and the constraint on $\gamma$ is largely prior-dominated, making this parameter basically unconstrained by this analysis.
Similarly, for the mass scales, higher $\Xi_0$ gives lower inferred redshift at given observed $\dgw$, so higher inferred source-frame mass. 
In App.~\ref{sec:SNRcuts} we further discuss the dependence of the results on the choice of threshold in the SNR, while in App.~\ref{sec:mmax} we discuss the dependence on the choice of the prior on 
$m_{\rm max}$.

\section{\label{sec:forecasts}Forecasts  and impact on population analysis}

\begin{table*}[th]\label{tab:fiducials}
 \begin{tabular}{@{}cccccccccccccccccccccccc@{}}
 \toprule
  \textbf{Parameter} & $H_0$ & $\oma$   & $R_0$ & $\gamma$ & $k$ & $z_p$   & $\alpha_1$ & $\alpha_2$ & $\beta$ &  $\delta_m$ & $m_{\rm min}$& $m_{\rm max}$ & $b$  \\ 
  \hline
   \textbf{Fiducial value} & 67.74 & 0.31    &    50 & 3 & 2 & 2   & 1.6 & 5.4 & 1.4 &  5 & 4     &     70 & 0.5  \\ 
  \botrule
 \end{tabular}
 \caption{Summary of fiducial values used for generating mock observations. 
 $R_0$ has units of $\text{Gpc}^{-3} \text{yr}^{-1}$, $m_{\rm min}$, $m_{\rm max}$  have units of $\msun$, and $H_0$ is in ${\rm km}\, {\rm s}^{-1}\, {\rm Mpc}^{-1}$.
With these choices, the fiducial value for $m_{\rm break}$ is $m_{\rm break}= 37 \, M_{\odot}$.
For modified GW propagation, we consider different values of $\Xi_0$ in Fig.~\ref{fig:Ndet} and we perform the full inference for the two cases $\Xi_0=1$ and $(\Xi_0=1.8, n=1.91)$.}
\label{tab:fiducials}
\end{table*}

The results presented in Sect.~\ref{sect:fullcosmopop} provide a joint constraints on modified GW propagation and population parameters together, using current data, but these are of course quite broad due to the limited statistics. However, this shows the potential for this method to constrain $\Xi_0$, and some interesting correlations with this parameter already show up in the data. Motivated by this, in this section we study the constraining power of five years of observations with an advanced LIGO/Virgo detector, both to understand to which extent we could constrain modified GW propagation, and to display the impact of neglecting this phenomenon while doing population studies.

We draw mock events from a population distribution described by a BBH mass function and a rate $R(z)$ with the functional forms given in sect.~\ref{sec:method}.
We consider different fiducial values for $\Xi_0$ (see below) while the fiducial values of the remaining parameters are given in Table~\ref{tab:fiducials}.
 As a detection criterion 
we require that the events have a $\rm SNR \geq 8$ in a single advanced LIGO detector with waveforms computed with the $\tt{IMRPhenomXAS}$ approximant and a PSD given by the $\tt{aLIGODesignSensitivityP1200087}$ available in $\tt{pycbc}$~\cite{alex_nitz_2020_3697109}. The uncertainty on the measurement is modeled as in Ref.~\cite{Farr:2019twy} (see also \cite{Fishbach:2018edt} for a detailed description and \cite{Farr19Git} for a python implementation). We draw samples from the likelihood of the simulated events as described in~\cite{Farr:2019twy}, thus simulating an actual observational situation and being able to analyze the dataset with the same code used in the actual analysis of GWTC-3.
 In order to compute the selection effects as described in App.~\ref{sec:SelBias}, we also generate a set of injections with a fixed reference population, and use for them the same detection criterion of the mock observations.

\begin{figure}[t]
\centering
\includegraphics[width=0.5\textwidth]{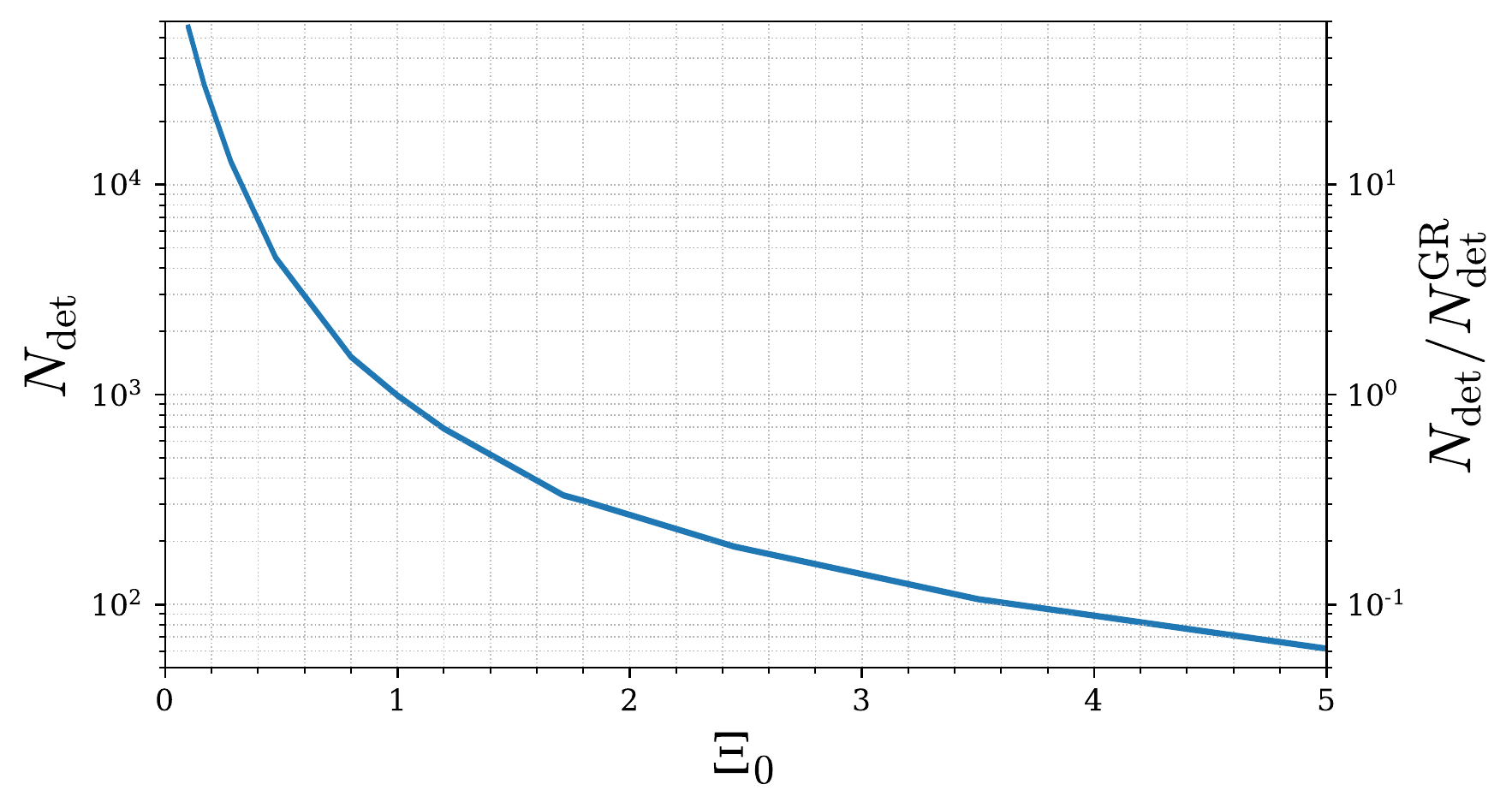}
\includegraphics[width=0.5\textwidth]{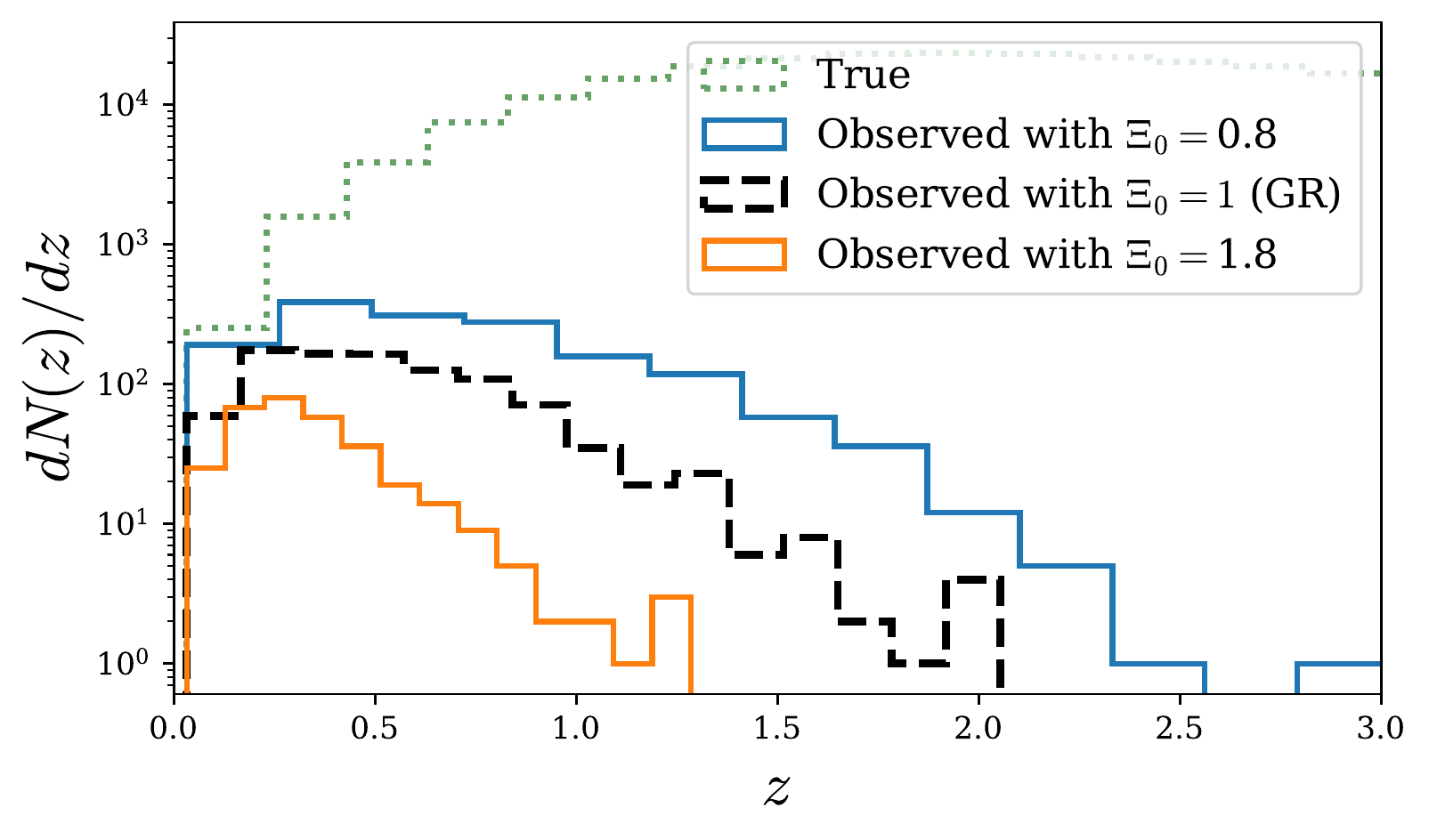}
\caption{Top: the expected number of detections in one year of observations by advanced LIGO/Virgo for the population model described in the text, as a function of $\Xi_0$. On the right vertical axis the ratio to the detections in GR ($\Xi_0=1$) is shown. Bottom: the differential number of detected events in one year of observations of advanced LIGO/Virgo, $dN(z)/dz$, as a function of redshift, for different values of $\Xi_0$. The original distribution is also shown (green, dashed line). }
\label{fig:Ndet}
\end{figure}

A first important result  is that the number of detections in a given time span has a very strong dependence on $\Xi_0$, as we see from Fig.~\ref{fig:Ndet}, where we show the number of detections in one year of observations as a function of $\Xi_0$ (top panel), and the distribution in redshift of the observed events for different values of $\Xi_0$ (bottom panel). We note that the effect can be quite relevant, giving already $\simeq 30\%$ less events for $\Xi_0\simeq 1.2$, $\simeq 70\%$ less events for $\Xi_0\simeq 1.8$, or, on the opposite side, $\simeq 30\%$ more events  for $\Xi_0\simeq 0.6$.
As we already discussed, the reason is that, for $\Xi_0>1$, events at a given redshift result in a GW luminosity distance which is larger than it would be in GR; then, these events give a lower SNR and are more difficult to detect. Hence, for a given merger rate density, the number of detected events is lower than in GR. Conversely, for $\Xi_0<1$ the detection rate is higher. Of course, the effect is much larger on the high-redshift tail of the distribution (bottom panel of Fig.~\ref{fig:Ndet}).
We will then present the results for two different choices of the fiducial values of $\Xi_0$,  the GR value $\Xi_0=1$ (in which case the other parameter $n$ that enters in \eq{eq:fit} is irrelevant), and the value  $\Xi_0=1.8$ (in which case we set $n=1.91$) inspired by the   prediction of  the RT nonlocal gravity model, which gives currently the largest deviation from GR for a viable and predictive model, see \cite{Belgacem:2020pdz} and footnote~\ref{note:inflation}. 
The other fiducial values are summarized in Tab.~\ref{tab:fiducials}.

\begin{figure}[t]
\centering
\includegraphics[width=0.5\textwidth]{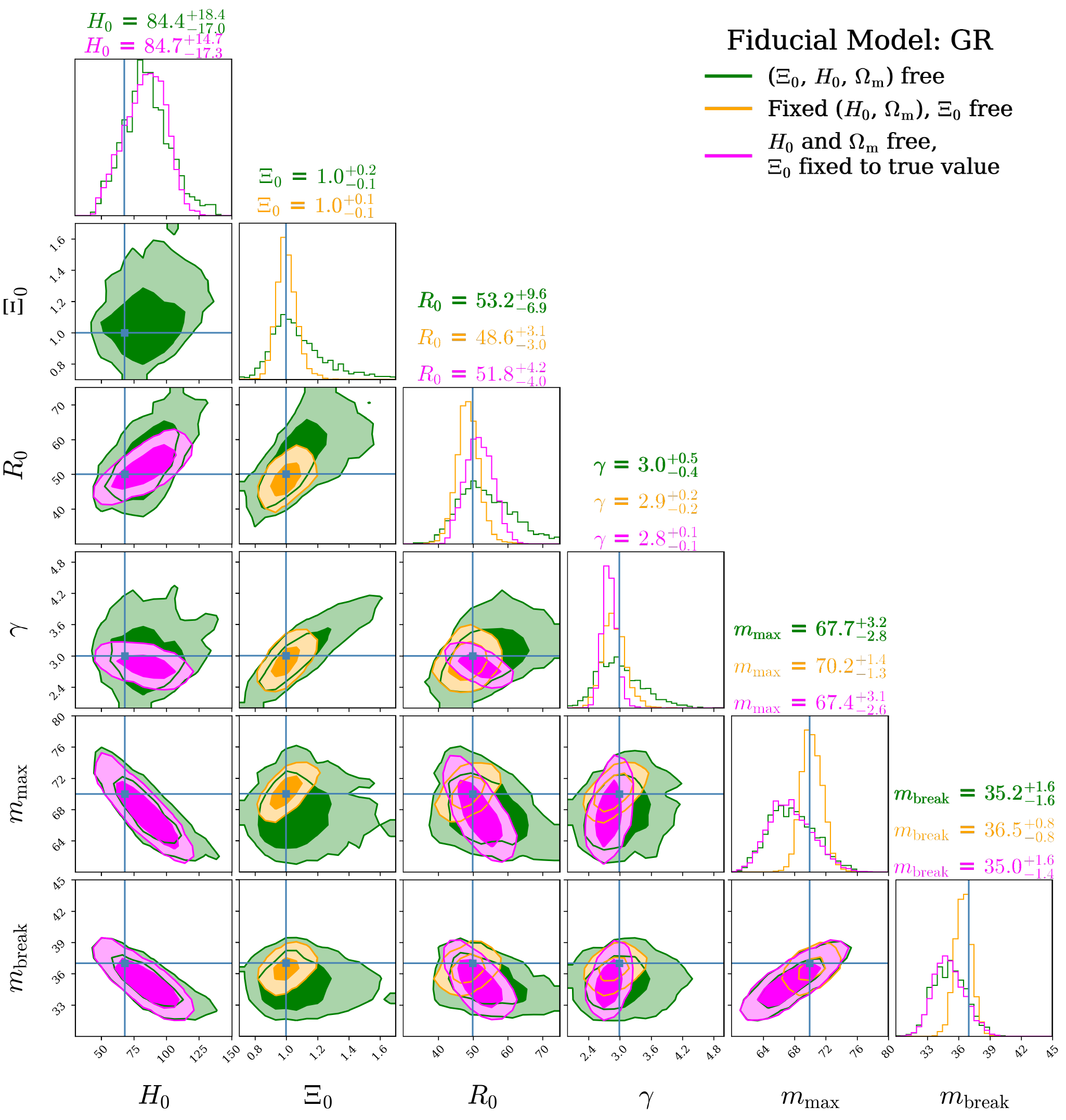}
\caption{Forecast for constraints on $ \{H_0, \Xi_0, R_0, \gamma, m_{\rm max}, m_{\rm break} \}$ with 5 years of observations of advanced LIGO/Virgo, for the fiducial value $\Xi_0=1$. The median and $68\%$ C.I. are reported, while the contours show the $68\%$ and $90\%$ confidence intervals. Lines denote the fiducial values. The orange contours correspond to fixing $H_0$ and $\oma$ to their true values, the green use a flat prior on both, and the purple ones further fix $\Xi_0=1$, mimicking a situation where GR is correctly assumed in the analysis.}
\label{fig:ForecastcornerGR}
\end{figure}

\subsection{GR as fiducial model}

We begin by discussing the case in which the fiducial value is  $\Xi_0=1$, in order to compare 
with the results of \cite{Farr:2019twy}. We  obtain $\sim 4700$ observations in 5 years, with a 100$\%$ duty cycle. This is comparable to the $\sim 5000$ events of Ref. \cite{Farr:2019twy}.\footnote{Even when setting $\Xi_0=1$, there are other differences with Ref.~\cite{Farr:2019twy}. First, the BBH mass function that we use in our work, following \cite{LIGOScientific:2020kqk}, assumes that the broken power law distribution for the primary mass describes the \textit{marginal} source-frame mass distribution $p(m_1)$, while in Ref.~\cite{Farr:2019twy} it is used to model the joint mass distribution $p(m_1, m_2)$. The marginal distribution $p(m_1)$ obtained in Ref.~\cite{Farr:2019twy} is shown in their Fig.~5; it can be seen that this model predicts many more masses in the range $30-40 M_{\odot}$, where they are more easily detected. This results in a much larger number of detected events. Second, the value  $R_0=60\, \text{Gpc}^{-3} \text{yr}^{-1}$ is used in Ref.~\cite{Farr:2019twy}. Finally,  the cutoff at the lower edge of the PISN mass gap, as modeled in Ref.~\cite{Farr:2019twy}, is sharper. Note, however, that a duty cycle of 50$\%$ is used in Ref.~\cite{Farr:2019twy}. 
Finally, when comparing the exact number of detections obtained when generating the mock datasets, one should keep in mind that some randomness is always present in the process, in particular a Poisson distribution for the number of events. \label{comparisonFarr}}
We consider three cases:  

\begin{enumerate} 

\item[(1).] The data are  analyzed fixing $\Xi_0$ to the same value used to generate the events, in this case the GR value $\Xi_0=1$, and we infer $H_0$ and $\oma$ from the data.

\item[(2).]   $\Xi_0, \oma$ and  $H_0$  are inferred simultaneously from the data.

\item[(3).]   $H_0$ and $\oma$ are fixed to the  values used to generate the events, and we only infer $\Xi_0$ from the data.

\end{enumerate}

The results are shown in Fig.~\ref{fig:ForecastcornerGR}.
Consider first the case (1).  Using a flat prior on both $H_0$ and $\oma$  the results are given by  the purple contours in Fig.~\ref{fig:ForecastcornerGR}. We  obtain
an accuracy on $H_0$ at $68\%$ c.l. of 
\be
\frac{\Delta H_0}{H_0} \simeq 20\%\, \quad   \text{(Fiducial: GR; $\Xi_0$ fixed)} \, ,
\ee
and an accuracy on $H(z_*)$ of about $7\%$ at a pivot redshift $z_*\simeq0.8 $.\footnote{The pivot redshift is defined as the value $z_*$ where the fractional uncertainty on $H(z)$ is minimised.} 
Comparing with Ref. \cite{Farr:2019twy}, we see that the pivot redshift remains consistent, while our prediction is more than a factor of two worse than the $\simeq3\%$ fractional uncertainty forecasted there (at the same pivot redshift), and also that, according to this updated forecast, such result would be achieved in twice the time, since we assumed a $100\%$ duty cycle instead of the $50\%$ used in Ref. \cite{Farr:2019twy}. Also, even if the accuracy on $H(z_*)$ remains below $\sim10\%$, the one on $H_0$ is still quite large. 
One origin for these differences is the different mass function used, as discussed in footnote~\ref{comparisonFarr}. We also point out  that, as a check,  we  run the inference with the parameters $k$, $z_p$ fixed to their fiducial values, in order to mimic a situation where the model for the merger rate is assumed to be simpler, as it is the case if one uses the  model $R(z) = R_0 (1+z)^{\kappa}$. We find that this leads to underestimating the error on $H_0$ by a factor of 2.  

Next, we include $\Xi_0$ in the inference while also letting $H_0$ and $\oma$  vary, case (2) above. This corresponds to the green contours in Fig.~\ref{fig:ForecastcornerGR}. In this case for $\Xi_0$ we get $\Xi_0=1.1^{+0.2}_{-0.1}$. 
For this fiducial, our forecast for the relative error on $\Xi_0$, with five years of LVC data, is then
\bees
\frac{\Delta\Xi_0}{\Xi_0}&\simeq& 14\%\,\\
&& \,   \text{(Fiducial: GR; marginalised over $H_0,\oma$)}.\nn
\ees
For $H_0$ we obtain an accuracy of $\simeq 20\%$,
and an accuracy on $H(z_*)$ of $\simeq 15\%$ at a pivot redshift $z_* \simeq 0.6$. The less stringent limit on $H(z)$, compared to the analysis done with $\Xi_0$ fixed, is due to the fact that the inclusion of $(\Xi_0, n)$ as free parameters affects the reconstruction of the entire function $\dgw(z)$, and in turn of $H(z)$, and cumulates with redshift. So, $H(z)$ is more affected at high redshift, while the accuracy on $H_0$ remains comparable since the effect of modified GW propagation is subdominant at very low redshift.

Finally, we consider the case (3), where $H_0$ and $\oma$ are assumed to be known exactly, which is an approximation to the situation where they are allowed to vary only within a prior range, fixed by external datasets (such as Cosmic Microwave Background, Baryon Acoustic Oscillations, and supernovae), quite narrow compared to that of the other parameters. The result is shown by the orange contours in Fig.~\ref{fig:ForecastcornerGR}. In this case for $\Xi_0$ we get $\Xi_0=1.0^{+0.1}_{-0.1}$, corresponding to a fractional uncertainty 
\be
\frac{\Delta\Xi_0}{\Xi_0}\simeq 10\%\, \,  \quad \text{(Fiducial: GR; fixing $H_0$, $\oma$)}.
\ee 
It is also interesting to observe, again from Fig.~\ref{fig:ForecastcornerGR} with fixed $H_0$ and $\oma$, that with five years of LVC data we forecast an accuracy of  about  $2 \%$ on $m_{\rm break}$,  $2 \%$ on $\mmax$, and $6 \%$ on $R_0$. Another point of interest for these parameters is that marginalising over $\Xi_0$ does not affect the accuracy on the mass scales with respect to the case where $\Xi_0$ is fixed (green and magenta contours respectively). On the contrary, the accuracy on the parameters $R_0$ and $\gamma$ describing the merger rate is degraded by about a factor of 2 if $\Xi_0, H_0, \oma$ are inferred simultaneously. This is consistent with what we observed in the actual analysis of GWTC-3, where we found that the parameter $\gamma$ is largely unconstrained if $(\Xi_0,n)$ are let to vary, differently from $m_{\rm break}$.

\subsection{Modified gravity as fiducial model}

We now set $\Xi_0=1.8$ as fiducial value. We consider again  three scenarios: one in which the values of $H_0$ and $\oma$ are known, one in which we impose a flat prior over both - to investigate the constraining power on $H_0$ and $\Xi_0$ jointly - and one in which, in the inference, $\Xi_0$ is wrongly fixed to the GR value 
$\Xi_0=1$ while the mock data  have been generated using $\Xi_0=1.8$; this allows us to study the bias that would be induced on the population parameters if Nature is described by a modified gravity theory which predicts modified GW propagation with such a large difference from GR, and the data are analysed ignoring this phenomenon. 
\begin{figure}[t]
\centering
\includegraphics[width=0.5\textwidth]{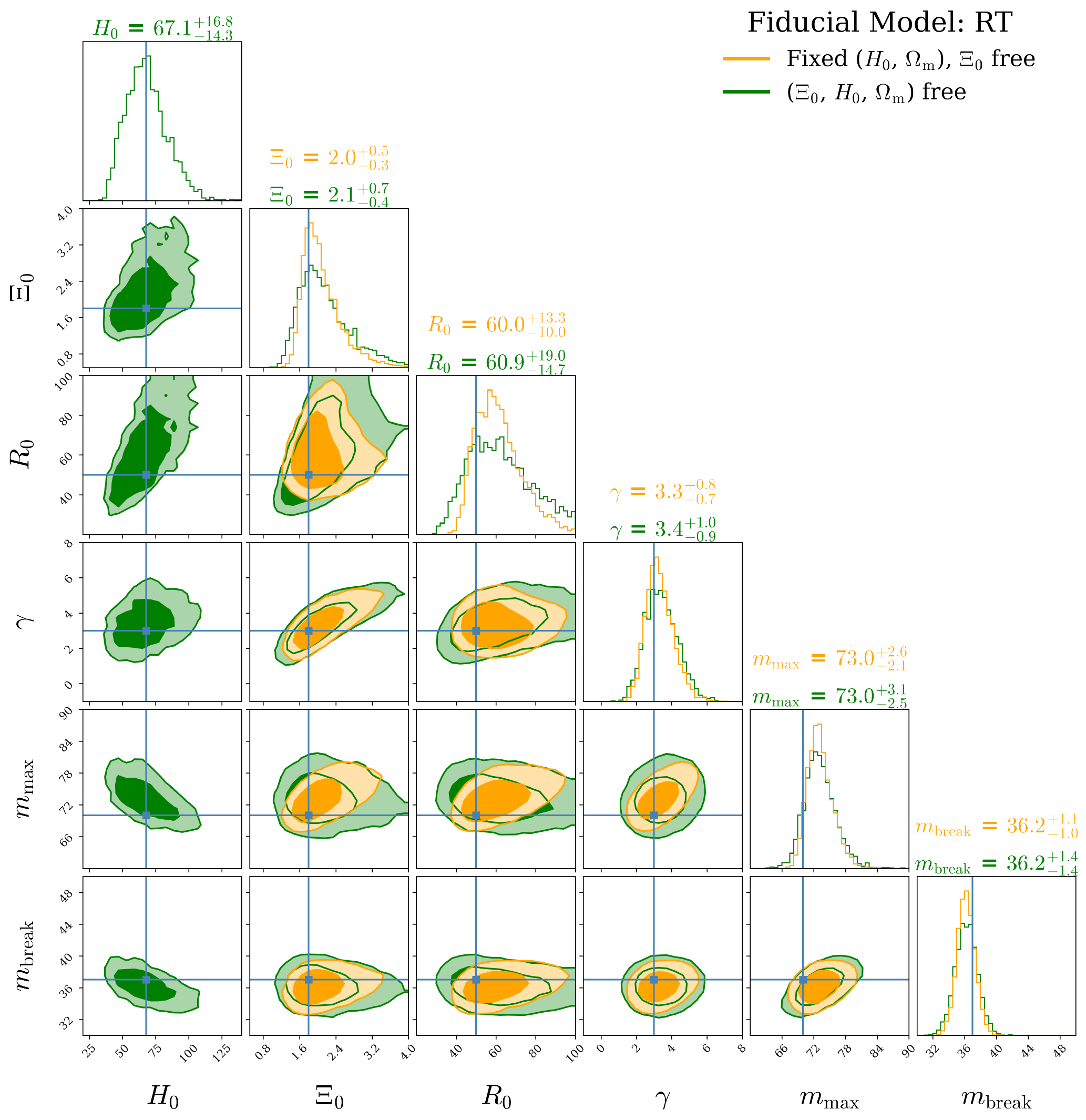}
\caption{Forecast for constraints on $ \{H_0, \Xi_0, R_0, \gamma, m_{\rm max}, m_{\rm break} \}$, with 5 years of observations of advanced LIGO/Virgo, when the fiducial model is a modified gravity model with $\Xi_0=1.8, n=1.91$, as in the RT nonlocal model. The median and $68\%$ C.I. are reported, while the contours show the $68\%$ and $90\%$ confidence intervals. Lines denote the fiducial values. The orange contours correspond to fixing $H_0$ and $\oma$ to their true values, while the green use a flat prior on both. }
\label{fig:Forecastcorner}
\end{figure}

The results for the forecast for five years of LIGO/Virgo data are shown in Fig.~\ref{fig:Forecastcorner}, in the two cases of fixed value of $H_0, \oma$ (orange) and for a flat prior on both (green). In particular we observe that, with $H_0$ fixed, the fiducial value $\Xi_0=1.8$ is correctly recovered, with a $68\%$ C.I. $\Xi_0=2.0^{+0.5}_{-0.3}$, that would already provide a hint for a deviation from GR, which lies at $\approx 2.5 \sigma$, though not yet a compelling evidence. 
For this fiducial, our forecast for the relative error on $\Xi_0$, with five years of LVC data,  is then
\be
\frac{\Delta\Xi_0}{\Xi_0}\simeq 20\%\,  \quad \text{(Fiducial: RT; fixing $H_0$, $\oma$)}.
\ee 
Without fixing $H_0$, we obtain instead a $26\%$ accuracy.

For $H_0$, using a flat prior, we get 
an accuracy at the level of $23\%$, comparable to the accuracy for $\Xi_0$. The result for $\Xi_0$ is, however, significantly worse than that obtained using $\Xi_0=1$ as a fiducial. This is due to the fact that with the choice $\Xi_0=1.8$ as fiducial value, the GW luminosity distance for a given true redshift is increased with respect to the GR case. Hence, events at higher redshift, that are the ones useful to constrain $\Xi_0$, easily fall out of the detector horizon. The overall number of detected events is actually significantly lower, and the accuracy scales roughly as $\sim1/\sqrt{N_{\rm det}}$: for  $\Xi_0=1.8$ we get in fact $\sim1600$ events, compared to the $\sim4700$ events that we got for $\Xi_0=1$. On the other hand, $H_0$ is constrained by events at low redshift, that have more chances of remaining detectable even increasing the fiducial value of $\Xi_0$. Hence, the corresponding accuracy is less degraded with respect to the GR case, because the ``effective number'' of events contributing to the constraint on $H_0$ remains  comparable, although a number of them is observed with larger errors.  This is coherent with the lower panel of Fig.~\ref{fig:Ndet}, where we see that,  at redshifts $\lesssim 0.5$,  the number of detected events for $\Xi_0=1.8$ is  of the same order of magnitude to the GR case. For higher redshifts, in contrast, the detection rate for $\Xi_0=1.8$ drops dramatically.

Finally, for this fiducial value of $\Xi_0$, with five years of LVC data we forecast an accuracy of  about  $3\%$ on $m_{\rm break}$,  $3\%$ on $\mmax$, and $19\%$ on $R_0$.
\begin{figure}[t]
\centering
\includegraphics[width=0.5\textwidth]{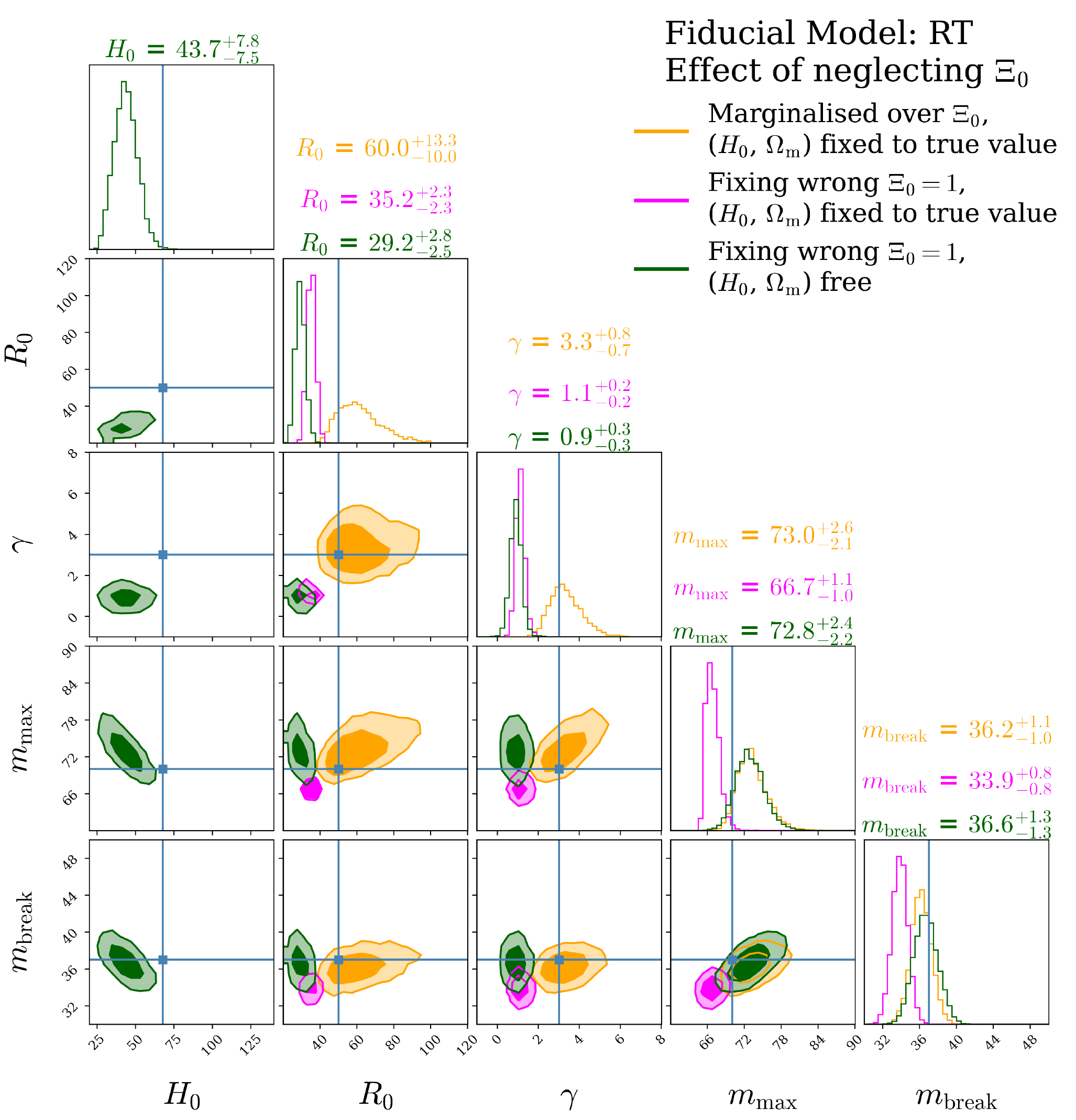}
\caption{Impact of neglecting the effect of modified GW propagation on the inference of population parameters. The magenta contours correspond to an analysis with $\Xi_0$ fixed at the GR value of 1 (with the true value being $1.8$) while $\oma$ and $H_0$ are fixed at the true values; the green contours correspond to fixing $\Xi_0$ at the GR value of 1, but letting $\oma$ and $H_0$ vary (with a flat prior on both); the orange contours have been marginalized over $\Xi_0$ with $\oma$ and $H_0$ fixed at the true values, i.e. they correspond to the orange contours in Fig.~\ref{fig:Forecastcorner}. When wrongly fixing $\Xi_0$ to the GR value, a visible bias appears especially in the rate evolution parameter $\gamma$. The median and $68\%$ C.I. are reported, while the contours show the $68\%$ and $90\%$ confidence intervals. Lines denote the fiducial values. }
\label{fig:bias}
\end{figure}
\\

As the last case, we  study the bias that would be induced on the estimate of the cosmological parameters, if Nature is described by a modified gravity theory with a value of $\Xi_0$ that differs significantly from the GR value $\Xi_0=1$, and the data are erroneously interpreted in the context of GR.  
To do so, we consider again the dataset obtained with the fiducial values of Tab.~\ref{tab:fiducials} and $\Xi_0=1.8$, but during the inference we `wrongly' fix $\Xi_0=1$. We consider again separately the scenarios where $H_0$ and $\oma$ are  known or unknown.
The results are shown in Fig.~\ref{fig:bias}. When $H_0$ and $\oma$ are fixed to their true values (magenta contours), we see that significant biases appear in all variables. In particular, for this choice of a fiducial value $\Xi_0>1$, the inferred rate is lower than the actual one, coherently with the fact that the number of detected events for $\Xi_0=1.8$ is significantly lower than in the case $\Xi_0=1$ (see Fig.~\ref{fig:Ndet}).

As for the bias in the mass scales, this can be explained with the fact that, for a fiducial cosmology with $\Xi_0>1$,  the true values of all mass scales are higher than those  inferred by wrongly assuming GR, as we discussed in sect.~\ref{sec:method}. More quantitatively, 
from \eq{mdet_vs_ms}, for each mass in each event  
\bees\label{msource_true}
m_{\rm true}&=&\frac{m^{{(\rm det)}} }{1+z_{\rm true} } = 
\frac{m^{{(\rm det)}} }{1+z_{\rm GR} }\, \frac{1+z_{\rm GR} }{ 1+z_{\rm true} }\nn\\
&=& m_{\rm GR} \, \(\frac{1+z_{\rm GR} }{ 1+z_{\rm true} }\)\, ,
\ees
where $m_{\rm true}$ is the true value of the source frame mass, $z_{\rm true}$ is the true redshift of the source, $z_{\rm GR}$ is the redshift that would be incorrectly inferred from $\dgw$ assuming GR when Nature is described by a modified gravity theory with  $\Xi_0\neq 1$, and $m_{{\rm GR} }$ is the source-frame mass that would be incorrectly inferred using GR.
For example, consider an event with primary mass just at the edge of the gap for this model, $m_{1, \rm true}=70 M_{\odot}$, at a redshift  $z_{\rm true}=0.35$ (which roughly corresponds to the peak of the distribution in redshift of the detected events for our mock dataset). If the correct value in Nature is $\Xi_0=1.8$, the  GW luminosity distance is $\dgw\simeq2.58\, \rm Gpc$. Inferring the redshift of the source from this measured value assuming GR, one would get $z_{\rm GR}\simeq0.45$, and, inverting  \eq{msource_true}, $m_{\rm GR} \simeq65.1 M_{\odot}$, in agreement with the median value of $m_{\rm max}$ inferred from the actual MCMC analysis (magenta contours in Fig.~\ref{fig:bias}).

Interestingly, when $\oma$ and $H_0$ are allowed to vary (green contours in Fig.~\ref{fig:bias}), we find that the mass scales are instead recovered without bias, at the price of a bias in the inferred value of $H_0$. In particular, in this case the value of $H_0$ is lower than the fiducial. This is explained by the fact that, fixing $\Xi_0$ to a value lower than the true one, the inferred redshift at given GW luminosity distance increases with respect to the true one, and source frame mass scales are pushed to lower values, as in the explicit example discussed just above. Lowering $H_0$ has the opposite effect of shifting the inferred mass scales towards higher values again, compensating the bias. In turn, this leaves also a bias in the inferred merger rate parameters, which are again lower than the fiducial values, balancing the fact that with $\Xi_0=1$ one would expect much more events than with $\Xi_0=1.8$. This cannot be completely compensated by a shift in $H_0$, being a much stronger effect.

\section{\label{sec:Conclusion}Conclusions}

In this paper, following the general strategy proposed in  \cite{Farr:2019twy}, we have performed a joint hierarchical Bayesian analysis of the cosmological parameters, including modified GW propagation, and of the astrophysical parameters that describe  the BBH merger rate and the BBH mass function, using the GWTC-3 catalog of detections. In particular, the mass scales that enter the BBH mass function allow  one to break the degeneracy between source-frame masses and redshift, and obtain statistical information on the redshift of the observed events, and therefore, eventually, to extract information both on the cosmological  parameters and on the population parameters. 

Our results for the parameter $\Xi_0$ that describes modified GW propagation are summarized in  Table~\ref{tab:constraintsXi0}, using a  prior flat in $\Xi_0$ or a prior flat in $\log\Xi_0$.
These are currently the most stringent limits on this parameter, with a relative  error, at $68\%$ C.I., of about  $60\%$.  The corresponding posteriors are shown in Fig.~\ref{fig:GWTC3Xi0FlatvsLog}. 
The difference in the result among these two choices of prior will eventually disappear as the statistics increases and the peak in the posterior become more and more narrow; with current statistics the two results are very well consistent, although still somewhat different.

We have shown that (with our choice of threshold in the signal-to-noise ratio, ${\rm SNR} > 12$), the result is 
robust with respect to  the inclusion or exclusion of the potential outlier GW190521.
In app.~\ref{sec:SNRcuts} and \ref{sec:mmax} we explored the dependence  on other choices, such as lower values of the  threshold in SNR, or a narrower  prior on $m_{\rm max}$.
Our result depends on the presence of the scale $m_{\rm break}$, which remains robust. In any case, modeling the population will remain the main source of potential systematics as the statistical errors decrease, and the possible presence of population outliers should be taken into account. 

Finally, we have investigated the correlation between the various hyper-parameters describing the cosmology and the BBH population, finding in particular significant correlations between $\Xi_0$ and the parameters $R_0$ and $\gamma$ that describe the BBH merger rate, see Fig.~\ref{fig:GWTC2corner}.

At the level of the GTWC-3 catalog, the constraints that can be obtained on $H_0$ and on $\Xi_0$, with this technique based on the BBH mass function, are still quite broad. We have then generated mock data simulating  five years of LVC data, and we have performed forecasts on the accuracy that will be obtained on the cosmological and population parameters with such data. First of all, we have found that the number of BBH events detected in a given time span is very sensitive to the value of $\Xi_0$. Values of $\Xi_0>1$ correspond to a stronger attenuation of the GW amplitude during the propagation, compared to GR, and therefore effectively decrease the detector range in redshift, while, conversely, $\Xi_0<1$ has the effect of increasing it. 
This strong effect on the rate is quite interesting in itself, and of course also influences the accuracy that can be obtained on the cosmological and population parameters, in particular those that are more constrained by events at high redshift, as in the case of $\Xi_0$, since it affects the size of the sample of detections obtained in a given observation time, making harder to observe the events at high redshift.

We have then focused on two cases, the fiducial value $\Xi_0=1$, corresponding to GR, and the fiducial value $\Xi_0=1.80$, that we have motivated from the RT nonlocal model. For the relative accuracies on $H_0$ and $\Xi_0$ we get
\be
\frac{\Delta H_0}{H_0}\simeq 20\%\, ,\qquad
\frac{\Delta\Xi_0}{\Xi_0}\simeq 10 \%\, ,\hspace{5mm} (\Xi_0=1)\, ,
\ee
and
\be
\frac{\Delta H_0}{H_0}\simeq 23\%\, ,\qquad
\frac{\Delta\Xi_0}{\Xi_0}\simeq 20\%\, ,\hspace{5mm} (\Xi_0=1.8)\, ,
\ee
These forecasts indicate  that future LVC observations will already be able to probe modified GW propagation at the level where viable modified gravity models predict that a signal could be found.
As for $H_0$, the forecasted accuracy remains quite large, suggesting that this method alone, based on the use of the BBH mass function, will  not be able to  provide a resolution of the Hubble tension with LVC data. Combination of this method with other statistical probes will be particularly beneficial, in particular with the correlation of GW events and galaxy catalogs. We also note that, with the fiducial model adopted in this study which is closer to the latest constraints and uses a realistic model for the merger rate evolution, also the forecasted accuracy for $H(z_*)$ at the pivot redshift $z_*\sim0.8$ is degraded by more than a factor 2 with respect to previous study in \cite{Farr:2019twy}, being around $7\%$ if the fiducial model is assumed to be GR. In any case, this still represents a $<10\%$ determination of $H(z)$ at high redshift, which remains a remarkable result to be achieved with GWs. 
We have finally studied how the estimate of population parameters would be affected, if interpreted in the context of GR, when the correct theory of Nature has a significantly different value of $\Xi_0$, taking again $\Xi_0=1.8$ as our default reference value. The results, shown in Fig.~\ref{fig:bias}, show that quite significant biases would be induced.

In this paper we restricted ourselves to the broken power law mass model, which parametrises a single feature in the mass distribution of the heavier black hole. While at the current level this is enough to conduct the analysis presented here, it is important to be aware that more complicated distributions are being investigated \cite{LIGOScientific:2021psn}, some of which even introduce more than one feature in the mass distribution. In general, less defined mass scales can lead to weaker constraints on the cosmological parameters. However, current observations are dominated by statistical uncertainty that make this effect subdominant. On the other hand, this effect should be investigated in more detail when it comes to forecasting the constraints that can be obtained with future experiments, where the statistical uncertainty will become lower and the effect of less sharp mass scales can become more relevant. In summary, it is clear that modelling the population will be the most relevant source of systematic in this method. We leave for future work a thorough investigation of the effects of more complex population models.
Related to this point, after this work was completed, Ref. \cite{Leyde:2022orh} appeared, which conducts a very similar analysis, and considers in particular different mass models. The resulting constraints on $\Xi_0$ remain largely compatible among themselves - coherently with the above discussion -, as well as with the results presented in this paper\footnote{ An important difference concerns the prior choice on the rate evolution parameter $\gamma$. Ref. \cite{Leyde:2022orh} adopts a quite narrow Gaussian prior on the latter, with mean at $\gamma=2.7$ and unit variance, motivated by the results of population analyses \cite{LIGOScientific:2021psn}. On the contrary, in this work we chose to remain agnostic to $\gamma$, since a point of the paper is precisely that its value is strongly correlated to that of  $\Xi_0$, hence the two should be constrained together. The prior choice on $\gamma$ of Ref. \cite{Leyde:2022orh} results in a lower median value for $\Xi_0$, in agreement with the direction of the correlation in the $( \Xi_0, \gamma )$ plane found in our analysis (see Fig.~\ref{fig:GWTC2corner}).  }.

In summary, the use of the BBH mass function to infer jointly cosmological and population parameters is a very promising technique, that could lead to very significant results for tests of modified gravity in the near future, even more if combined with other statistical methods.

\vspace{5mm}
{\bf Code availability}.
Together with this paper, we release the publicly available code $\tt{MGCosmoPop}$, which is available under open source license at \url{https://github.com/CosmoStatGW/MGCosmoPop}. 

\vspace{5mm}
{\bf Acknowledgments}. We thank Andreas Finke and Francesco Iacovelli for many useful discussions, and Stefano Foffa for help in computing selection effects.
Computations made use of the Yggdrasil cluster at the University of Geneva.
The work of MM and MM is supported by the  Swiss National Science Foundation and  by the SwissMap National Center for Competence in Research. 
This research has made use of data, software and/or web tools obtained from the Gravitational Wave Open Science Center (https://www.gw-openscience.org/), a service of LIGO Laboratory, the LIGO Scientific Collaboration and the Virgo Collaboration.
Cornerplots are generated using $\tt{corner}$ \cite{corner}.

\appendix

\section{Computation of selection bias}\label{sec:SelBias}
In this appendix we discuss the computation of the term $N_{\rm exp}(\Lambda)$ in 
\eq{likelihood}. This represents the number of \emph{expected} events as predicted by the model for a given choice of $\Lambda$ and correctly accounts for selection bias~\cite{Loredo:2004nn, Mandel:2018mve}. 
Explicitly, it is given by
\begin{equation}\label{Ndetfull}
N_{\rm exp}(\Lambda) = \int_{ f(\mathcal{D})>\text{th.}  } d\mathcal{D} d\theta
\,  \mathcal{L(\mathcal{D} | \theta )} \frac{dN}{d\theta} \, ,
\end{equation}
where $f(\mathcal{D})$ is the selection function of the experiment, such that only data with $f(\mathcal{D})$ larger than a given threshold are detected. In our case, this function is given by the network SNR of the three-detector LIGO-Virgo network.
The integral in \eq{Ndetfull} can be computed by Monte Carlo integration, by drawing mock events from a reference distribution and storing those that pass the selection threshold together with the values of the parameters $\theta$ and the probability density $p_{\rm draw}(\theta)$ of the distribution used to generate the events~\cite{Tiwari:2017ndi}. 
This allows us to compute $N_{\rm exp}(\Lambda)$ with the estimator~\cite{Tiwari:2017ndi,Farr_2019}:
\begin{equation}\label{Ndet}
    x \simeq \frac{1}{N_{\rm draw}} \sum_{i=1}^{N_{\rm det}}\frac{1}{p_{\rm draw}(\theta_i)} \frac{d {N}}{d \theta_i}(\Lambda)
\end{equation}
where $N_{\rm draw}$ is the total number of generated events and $N_{\rm det}$ is the number of detected ones . To account for the uncertainty in $x$ we follow closely the calculation of Ref.~\cite{Farr_2019}. 
With a large number of samples, the distribution of the quantity $x$ is Gaussian: $x\sim \mathcal{N}(\mu, \sigma)$. The estimator $\hat{\mu}$ of the mean is equal to the RHS of \eq{Ndet}, while the variance can be estimated as
\begin{equation}\label{sigmaSq}
    \hat{\sigma}^2 = \frac{1}{N^2_{\rm draw}} \sum_{i=1}^{N_{\rm det}}{\left(\frac{1}{p_{\rm draw}(\theta_i)} \frac{d {N}}{d \theta_i}(\Lambda)\right)}^2 - \frac{\hat{\mu}^2}{N_{\rm draw}} \, , 
\end{equation}
From the above equation, one can also define the \emph{effective} number of independent samples that contributed to the computation of $x$, as $N_{\rm eff} \equiv \hat{\mu}^2/\hat{\sigma}^2$. In the inference, we ensure that $N_{\rm eff}>4 N_{\rm obs}$ ($N_{\rm obs}$ being the number of observations), otherwise the sample is rejected~\cite{Farr_2019}.

One should then marginalize the posterior in \eq{posterior0} over the distribution of $x$. Here we slightly adapt the calculation of Ref.~\cite{Farr_2019} by imposing that the normal distribution is truncated at zero in order to avoid non-zero probability for negative values of $x$.
The posterior becomes
\begin{equation}
\begin{split}
  p(\Lambda | \mathcal{D} ) & \propto  \pi(\Lambda )\,  \prod_{i=1}^{N_{\rm obs}} \left\langle\frac{1}{\pi(\theta_i)}  \frac{d {N}}{d \theta_i}(\Lambda) \right\rangle_{\rm samples} \, \\
  &\times \int_0^{\infty} dx \frac{N(x; \hat{\mu}, \hat{\sigma})}{\Phi(0; \hat{\mu}, \hat{\sigma})} \text{e}^{-x}  \,,
  \end{split}
\end{equation}
where $\Phi(0; \mu, \sigma)$ is the Complementary Cumulative Distribution Function of a Gaussian with mean $\mu$ and standard deviation $\sigma$, and ensures the correct normalization of the truncated Gaussian distribution. Integrating, we get the posterior
\begin{equation}\label{posteriorFull}
\begin{split}
  p(\Lambda | \mathcal{D} )  & \propto \pi(\Lambda )\,  \text{e}^{-\hat{\mu}(\Lambda)}  \prod_{i=1}^{N_{\rm obs}} \left\langle\frac{1}{\pi(\theta_i)}  \frac{d {N}}{d \theta_i}(\Lambda) \right\rangle_{\rm samples} \, \\
  & \times \text{e}^{\hat{\sigma}^2/2}  \frac{\Phi(0;  \hat{\mu}-\hat{\sigma}^2, \hat{\sigma})}{\Phi(0;  \hat{\mu}, \hat{\sigma})} \, ,
  \end{split}
\end{equation}
with $\hat{\mu}$ and $\hat{\sigma}$ given by the RHS of \eq{Ndet} and \eq{sigmaSq} respectively.

When doing a cosmological inference, it is possible that a variation of $\Xi_0$ induces a large variation in the inferred source frame masses corresponding to given detector frame masses. We therefore need to have a set of injections that cover sufficiently the primary and secondary masses as well as the redshift ranges, and it is not possible to use the results of the injection campaign used by the LVC for their population analysis, since this was performed for an analysis at fixed cosmology. We generate a set of injections as follows. First, we draw events from a reference population with mass function given by a  ``truncated power law'', i.e. 
\be
p(m_1,m_2|\alpha, \beta) \propto p(m_1|\alpha) p(m_2 | m_1, \beta) \, ,
\ee 
where 
\be
p(m_1|\alpha) \propto m_1^{-\alpha}\, ,
\ee
\be 
p(m_2 | m_1, \beta) \propto m_2^{\beta}\,  ,
\ee
and merger rate evolution with redshift 
\be
R(z)=R_0 (1+z)^{\kappa}\, ,
\ee 
with $\alpha=1.1, \beta=0.75, R_0 = 20 \text{Gpc}^{-3} \text{yr}^{-1}, \kappa=4$, between redshift $0$ and $5$ and for $m_1 \in [2, 500] \msun$, $m_2<m_1$. We sample the angles drawing the source position in the sky $(\theta, \phi)$ as well as the time of the day of the event from uniform distributions and the inclination angle $\iota$ from a distribution flat in $\cos{\iota}$. Then, we compute the network SNR of each sampled event for a three detector network composed by the Virgo and the two LIGO detectors, taking into account their position on the Earth at the given time  and the correct orientation of the interferometers' axes. We use PSDs indicative of the O1-O2, O3a, and O3b for each observation period, as given in Figures 2 of \cite{LIGOScientific:2020ibl} for O2, \cite{LIGOScientific:2020ibl} for O3a, and \cite{LIGOScientific:2021djp} for O3b.\footnote{The corresponding data are available at \url{https://dcc.ligo.org/P1800374/public/} and \url{https://dcc.ligo.org/LIGO-P2000251/public} for O2 and O3a, while we compute the PSD with the $\tt{pycbc}$ software around the GPS time indicated in Fig. 2 of  \cite{LIGOScientific:2021djp} for O3b.}
Finally, we take into account the duty cycles of each detector during the O1-O2, O3a and O3b observing run accordingly to the sensitivity status for each run.\footnote{Available at  \url{https://www.gw-openscience.org/summary_pages/detector_status/O2/},  \url{https://www.gw-openscience.org/detector_status/O3a/}, \url{https://www.gw-openscience.org/detector_status/O3b/}} We use the $\tt{ImrPhenomPv2}$ waveform for O1-O2 and O3a, and the $\tt{ImrPhenomXPHM}$ waveform for O3b to match the ones used in the data releases when possible (this is the case for O1-O2 and O3b, while for O3a a mixture of waveforms was used; we checked that the effect of using different waveforms is small).
This procedure still contains some approximations. One is that we use only the average duty cycle of the detectors during the observing runs. This results in an average network duty factor that gives a total time with no interferometer active which is lower than the one reported in the sensitivity status pages (e.g. $\sim 1.5 \%$ in our simulation vs $\sim 3 \%$ in the sensitivity status report for O3b). This is due to the fact that, not knowing the exact detector status at each time, the best we can do is to ``shut down'' each detector randomly at a fraction of the detection time corresponding to that given in the sensitivity status reports. 
The second approximation is the use of a fixed PSD. Incorporating the actual noise of the detectors throughout all the observing runs requires significantly larger computational efforts. However, using a relatively high SNR threshold such as 12, used in our main analysis, prevents us from being too sensitive to fluctuations in the detectors' PSD. A related point is that our threshold is imposed only on the network SNR and does not include conditions on the False Alarm Rate. However, we note that, once the SNR threshold of 12 is used, all the events considered have a very small $\text{FAR}<2 \times 10^{-4} \text{yr}^{-1}$. We also checked that, even when using a lower SNR threshold of 8, the population analysis of GWTC-2 with a broken power-law mass function perfectly agrees with the result of \cite{LIGOScientific:2020kqk} despite all the differences discussed so far.\footnote{We compared with the posterior samples reported in the data release associated to the  LVC population analysis paper \cite{LIGOScientific:2020kqk}, available at \url{https://dcc.ligo.org/LIGO-P2000434/public}.} In particular, all the mass scales, relevant for this analysis, match perfectly, while the only parameter with a slight difference is the overall rate $R_0$, for which we get $R_0=12.1^{+5.5}_{-3.9}$ at $68\%$ C.L., while in \cite{LIGOScientific:2020kqk}  $R_0=19.1^{+16.2}_{-9.0}$ at $90\%$ C.L. is found. This slightly lower rate is consistent with the fact that we find a smaller fraction of the observational time where no interferometer is active. The two results remain in any case statistically indistinguishable. Finally, we find very good agreement also with the result for $H_0$ obtained in \cite{LIGOScientific:2021aug}, see App.~\ref{sec:H0}.

\section{BBH mass distribution}\label{sec:massfunction}
In this Appendix we report for completeness the definition of the mass distribution $p(m_1, m_2 | \Lambda_{\rm m})$ [see App.~(B) of \cite{LIGOScientific:2020kqk}]. We write 
\bees
p(m_1,m_2|\Lambda_m)&=&
\pi(m_1 | \alpha_1, \alpha_2,\delta_m, \mmin, \mmax, m_\text{break})\nn\\
&&\times \pi(q | \beta_q, m_1, \mmin, \delta_m)\, ,
\ees
where $q=m_2/m_1$ and $m_1\geq m_2$.
The distribution of the primary mass $m_1$  is given by
\bees
&&\pi(m_1 | \alpha_1, \alpha_2,\delta_m, \mmin, \mmax, m_\text{break})\\
&&   
 \propto
 \begin{cases}
        m_1^{-\alpha_1} S(m_1|\mmin,\delta_m) & \mmin < m_1 < m_\text{break} \\
        m_1^{-\alpha_2} S(m_1|\mmin,\delta_m) & m_\text{break} < m_1 < \mmax \\
        0 & \text{otherwise}\, ,
\end{cases}\nn
\ees
where the function $S(m_1|\mmin,\delta_m)$ is a smoothing function, which rises from 0 to 1 over the interval 
$(\mmin, \mmin+\delta_m)$, explicitly given by:
\bees
&&S(m_1|\mmin,\delta_m)\\
&&   
 =
 \begin{cases}
        0 & m<\mmin \\
         [f(m-\mmin, \delta_m)+1]^{-1} & \mmin \leq m < \mmin+\delta_m\\\
        1 & m \geq \mmin+\delta_m\, ,
\end{cases}\nn
\ees
with 
\be
f(m', \delta_m) = \exp{ \Bigg( \frac{ \delta_m}{m'}+ \frac{ \delta_m}{m'- \delta_m}  \Bigg)} \, .
\ee

This model therefore describes two different power-law behaviors for the primary mass, matched at a mass scale $m_\text{break}$.  The latter can be expressed in terms of a dimensionless parameter $b$ as 
\begin{align}
    m_\text{break} = \mmin +
    b(\mmax-\mmin) ,
\end{align}
The conditional distribution of the secondary mass is
\begin{align}
\label{eq:pq_smoothing}
\pi(q | \beta_q, m_1, \mmin, \delta_m) \propto q^{\beta_q} S(q m_1 \mid \mmin, \delta_m).
\end{align}
The set of hyperparameters describing the BBH mass function is therefore 
\be\label{Lambdam}
\Lambda_m=\{\alpha_1, \alpha_2, \beta_q, \delta_m, m_{\rm min}, m_{\rm max}, b \}
\ee

\section{Results for $H_0$}\label{sec:H0}

\begin{figure}[t]
\centering
\includegraphics[width=0.5\textwidth]{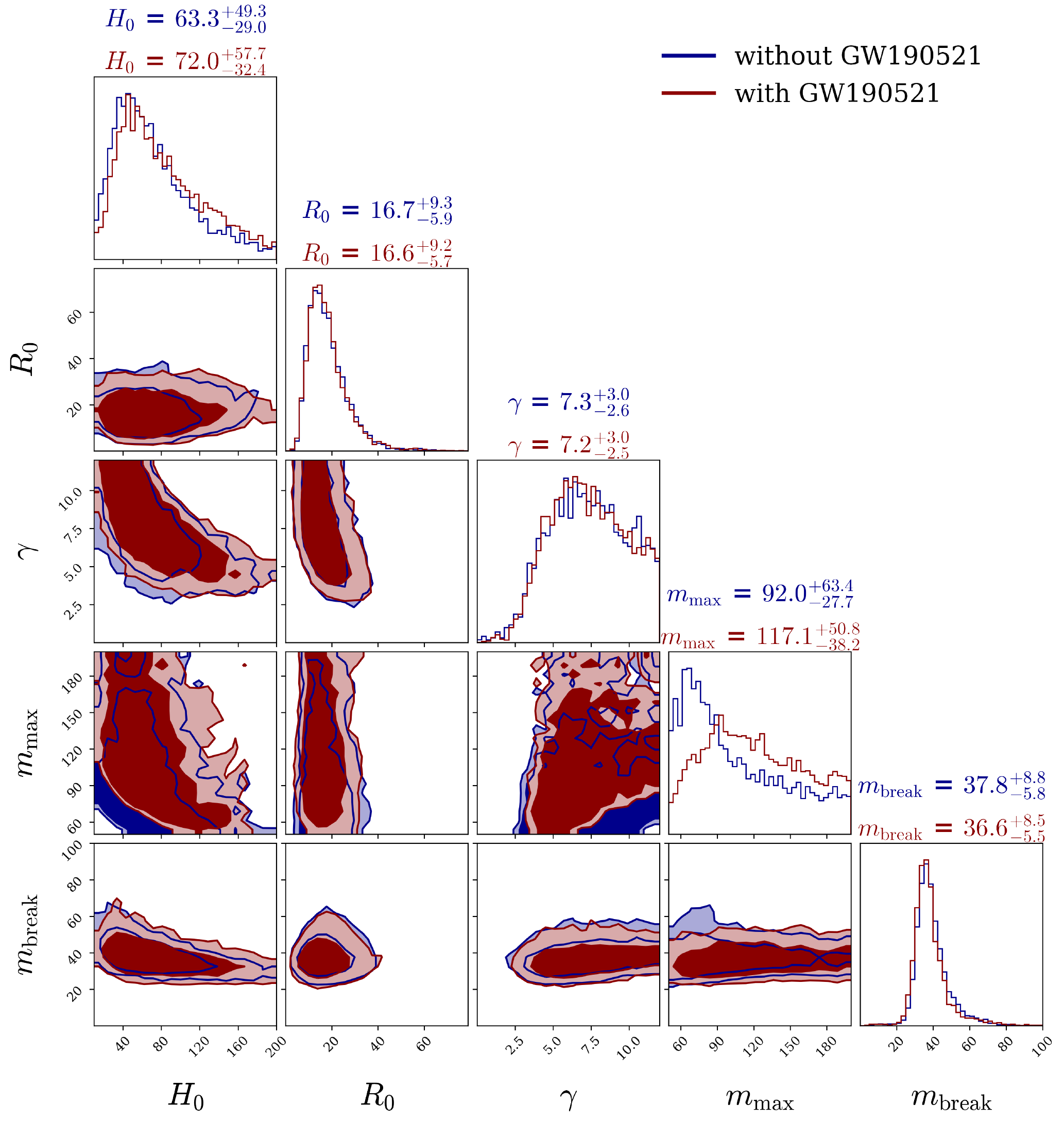}
\caption{Joint constraints from GWTC-3 on the parameters $ \{H_0, \oma, \gamma,m_{\rm max}, m_{\rm break} \}$, with $\Xi_0$ fixed to the GR value $\Xi_0=1$ and marginalized over all the other parameters. Contours show the $68\%$ and $90\%$ confidence intervals. The median and $68\%$ C.I. are reported.}
\label{fig:GWTC3cornerH0}
\end{figure}

In Fig.~\ref{fig:GWTC3cornerH0} we show results obtained by fixing $\Xi_0$ to the GR value $\Xi_0=1$ and adding $H_0, \oma$ as parameters in the inference, with flat priors between $[10, 200]$ and $[0.05, 1]$ respectively. We find, as in the case for $(\Xi_0, n)$ at fixed $H_0, \oma$, that the result is robust to the inclusion of GW190521 and depends on the mass scale $m_{\rm break}$, whose value is in perfect agreement with the one found in the main analysis. Our result is also in very good agreement with the recent one of~\cite{LIGOScientific:2021aug}. In particular, including GW190521 we find $H_0 = 72.0 ^{+93.9}_{-45.8}$ at $90\%$ C.L., in very good agreement with the value $H_0 = 69.0 ^{+88}_{-47}$ found in~\cite{LIGOScientific:2021aug}.

\section{Comparison of different SNR thresholds}\label{sec:SNRcuts}

In this appendix we examine the effect of using lower cuts on the SNR. We consider the two cases ${\rm SNR} > 11$ and ${\rm SNR} > 10$. In the latter case, we also exclude the event GW190424$\_$180648 which has a very high False Alarm Rate $\text{FAR}\simeq9 \, \text{yr}^{-1}$, and we obtain a sample of 41 events for the case ${\rm SNR} > 11$ and 53 events for ${\rm SNR} > 10$.
We do not consider lower SNR thresholds since this would lead to including more and more events with high FAR; given that the FAR is not included in the computation of the selection effects (see App.~\ref{sec:SelBias}), this could lead to a bias. For example, imposing a  cut  ${\rm SNR} > 8$, there remain $15$ events with $\text{FAR}>0.25 \, \text{yr}^{-1}$, among which 8 events have $\text{FAR}>2 \, \text{yr}^{-1}$. This start not only to be a relevant probability, but also to concern a non negligible fraction of the total number of events (85 events,  with ${\rm SNR} > 8$). 

\begin{figure}[t]
\centering
\includegraphics[width=0.4\textwidth]{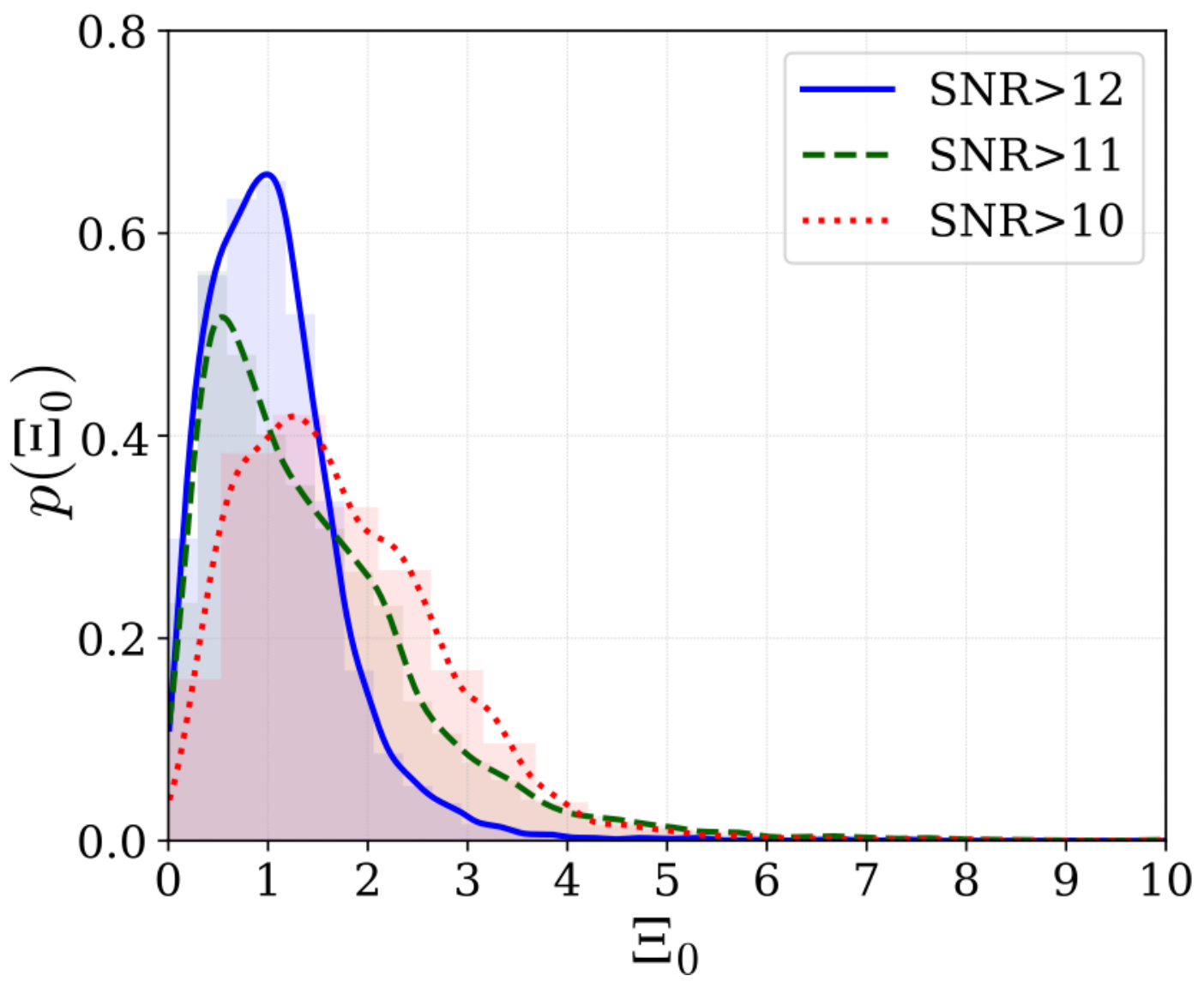}
\caption{Posterior probability for $\Xi_0$, marginalised over all the other parameters as in the main analysis, and for a prior flat in $\log\Xi_0$, with different SNR cuts.}
\label{fig:SBRsXi0}
\end{figure}

The results for the posterior   for  $\Xi_0$, marginalised over all the other parameters as in the main analysis,  and for a prior flat in $\log\Xi_0$, are shown in Fig.~\ref{fig:SBRsXi0}. They remain consistent within the statistical uncertainty, though we observe that lowering the SNR cut results in a longer tail of the distribution at large $\Xi_0$, likely due to the fact that events at higher luminosity distance are added as the threshold lowers, which have generally broader posterior distributions in mass and distance, since they are harder to measure. 
In particular, for the median and symmteric C.I., we find 
\bees\label{Xi0SNR}
\Xi_0 &=& 1.2^{+1.2}_{-0.7}\, \qquad ({\rm SNR } > 11, 68\%)\, , \\
\Xi_0 &=& 1.2^{+2.4}_{-1.0}\, \qquad ({\rm SNR } > 11, 90\%)\, , \\
\Xi_0 &=& 1.6^{+1.1}_{-0.8}\, \qquad ({\rm SNR } > 10, 68\%)\, , \\
\Xi_0 &=& 1.6^{+1.9}_{-1.1}\, \qquad ({\rm SNR } > 10, 90\%)\, ,
\ees
while for the maximum and HDI, 
\bees\label{Xi0SNR}
\Xi_0 &=& 0.5^{+1.4}_{-0.3}\, \qquad ({\rm SNR } > 11, 68\%)\, , \\
\Xi_0 &=& 0.5^{+2.4}_{-0.5}\, \qquad ({\rm SNR } > 11, 90\%)\, , \\
\Xi_0 &=& 1.3^{+1.0}_{-0.8}\, \qquad ({\rm SNR } > 10, 68\%)\, , \\
\Xi_0 &=& 1.3^{+1.9}_{-1.1}\, \qquad ({\rm SNR } > 10, 90\%)\, .
\ees

\section{Dependence on the prior choice for $m_{\rm max}$}\label{sec:mmax}

\begin{figure}[t]
\centering
\includegraphics[width=0.4\textwidth]{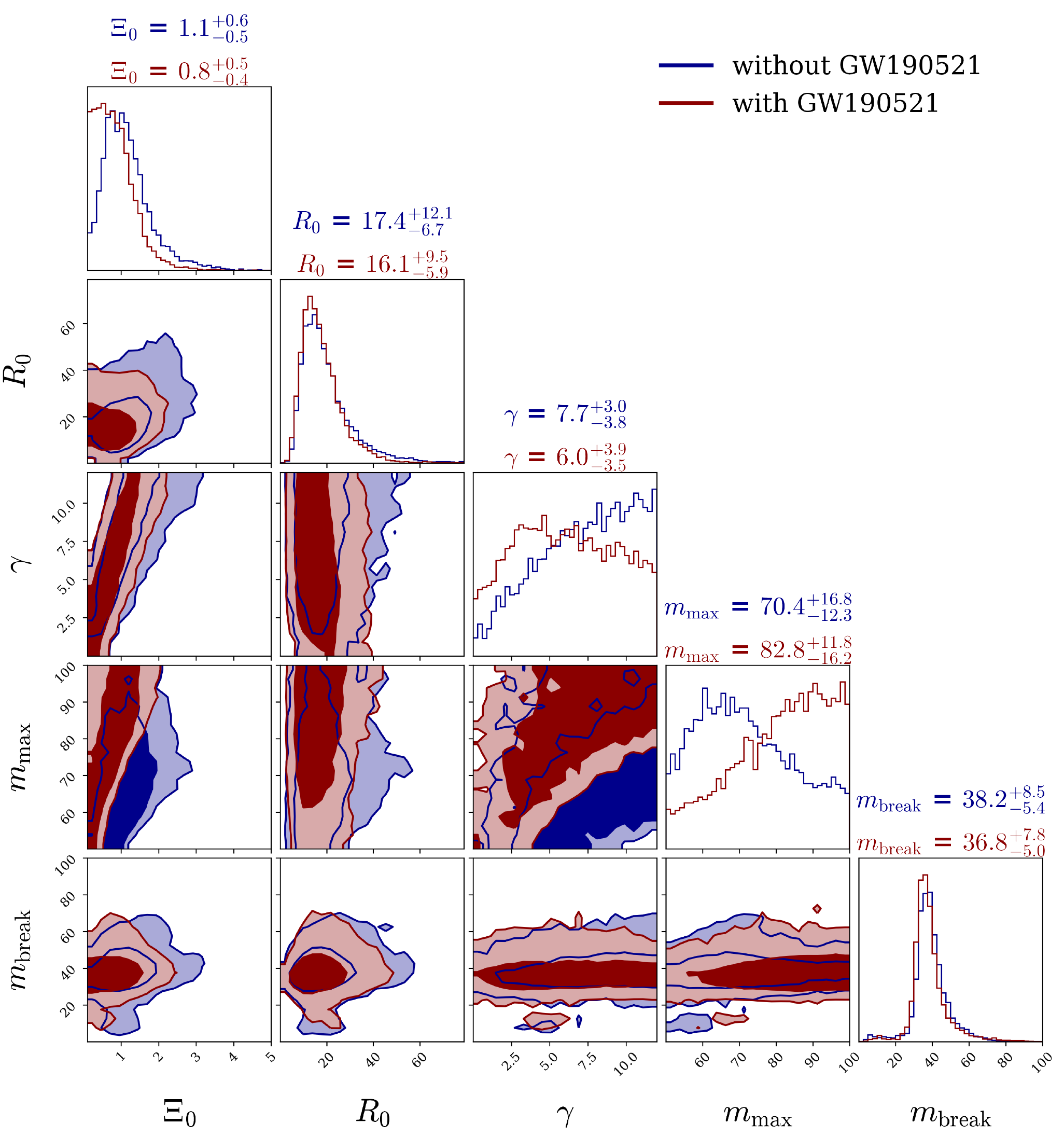}
\caption{Joint constraints from GWTC-3 on the parameters $ \{\Xi_0, R_0, \gamma,m_{\rm max}, m_{\rm break} \}$, for the sample with ${\rm SNR} > 12$ and with the same prior choices as in the main text, except for  $m_{\rm max}$ for which we now use a flat prior $\in [50, 100]\msun$. Contours show the $68\%$ and $90\%$ confidence intervals. The median and $68\%$ C.I. are reported.}
\label{fig:GWTC3cornerXi0mmax100snr12}
\end{figure}

\begin{figure}[t]
\centering
\includegraphics[width=0.4\textwidth]{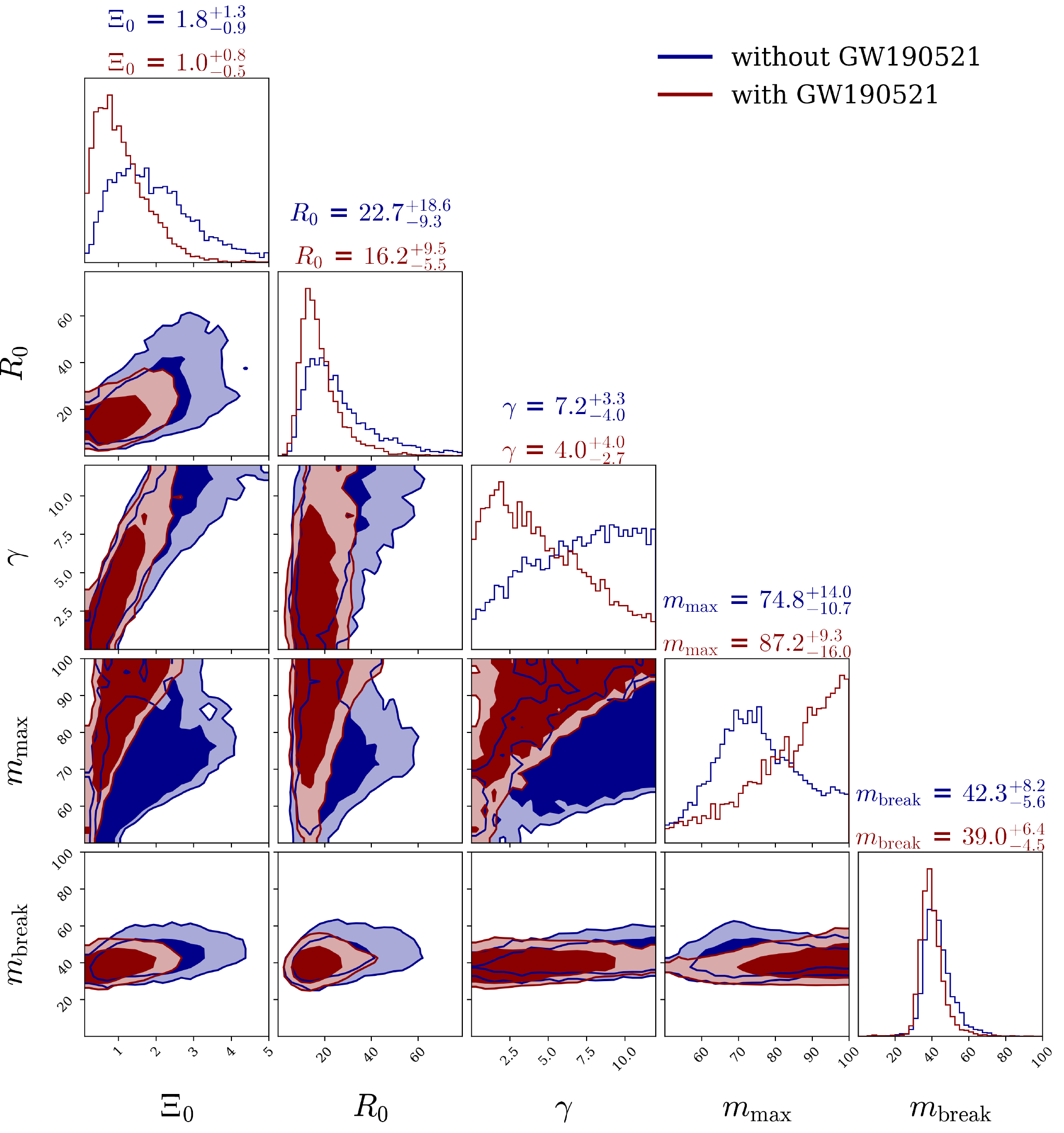}
\caption{Joint constraints from GWTC-3 on the parameters $ \{\Xi_0, R_0, \gamma,m_{\rm max}, m_{\rm break} \}$, for the sample with ${\rm SNR} > 10$ and with the same prior choices as in the main text, except for  $m_{\rm max}$ for which we now use a flat prior $\in [50, 100]\msun$. Contours show the $68\%$ and $90\%$ confidence intervals. The median and $68\%$ C.I. are reported.}
\label{fig:GWTC3cornerXi0mmax100snr10}
\end{figure}

As discussed in Sect.~\ref{sect:data}, the physical hypothesis that the BBHs come from a population featuring a mass gap can be translated in a hard cut on the upper limit of the prior for $m_{\rm max}$. However, this would make the analysis even more sensitive to the presence of population outliers at high mass. In particular, the event GW190521  has a significant fraction  (if not all) of  its posterior support for the primary mass within the gap, and a large part above $100\msun$, so its inclusion or not could become critical if using a hard prior on $m_{\rm max}$. In Fig.~\ref{fig:GWTC3cornerXi0mmax100snr12} we show the result of the analysis for the sample with ${\rm SNR} > 12$, using the same prior choices as in the main text, except for $m_{\rm max}$ for which we now use a flat prior $\in [50, 100]\msun$, while in Fig.~\ref{fig:GWTC3cornerXi0mmax100snr10} we show the result with the same restricted prior $m_{\rm max} \in [50, 100]\msun$ for the sample with ${\rm SNR} > 10$. 

We observe that, for this choice of prior, the sample with ${\rm SNR} > 12$ starts to show a small  difference in the marginalised posterior on $\Xi_0$, depending on the inclusion or not of GW190521.
In particular, including it in the analysis results in a posterior with more support for $\Xi_0<1$. This is easily explained, and is coherent with the discussion in Sect.~\ref{sect:data}: values of $\Xi_0<1$ have the effect of lowering the inferred source frame mass for a given detector-frame mass. Hence, imposing a hard cut for $\mmax$ has the effect of favoring low values of $\Xi_0$, in order to force the high mass of GW190521 within the prior range.  
Indeed, from Figs.~\ref{fig:GWTC3cornerXi0mmax100snr12} and \ref{fig:GWTC3cornerXi0mmax100snr10} we see that, when including GW190521, $\mmax$ squashes on the higher prior limit of $100\, M_{\odot}$ with a monotonically growing posterior until the upper limit of the prior is hit.  In contrast, if we exclude GW190521,  the result for $\Xi_0$ remains almost identical to the one obtained in the main analysis. 

As can be seen in Fig.~\ref{fig:GWTC3cornerXi0mmax100snr10}, the effect of GW190521 is enhanced when  the SNR threshold is lowered. In this case, the difference between the two posteriors for $\Xi_0$ is larger, and the effect of the hard prior forcing $\Xi_0$ to assume lower values is much more evident.
In this case, this also  has the effect of changing the shape of the posterior for the merger rate slope $\gamma$ (which in any case still remains prior dominated). The shift towards lower values is explained by the fact that, for given luminosity distance, the inferred redshift is higher for lower $\Xi_0$, hence $\gamma$ is forced to assume lower values.

In general, all the results remain compatible within the statistical uncertainties. However, the enhanced shift of the posterior for $\Xi_0$ in the case of lower SNR cut and narrow prior for $\mmax$ suggests that, the lower the SNR cut, the more the constraint on $\Xi_0$ is driven by the prior choice on $\mmax$ (which is highly sensitive to the presence of population outliers) rather than by the presence of a scale at $m_{\rm break}$, which remains robust. Further lowering the SNR cut may enhance this effect. Such a strong effect from a single event, in a sample of 53 events, might suggest that GW190521 is indeed an outlier, even with respect to the broken power-law mass distribution, but more statistics is of course needed to draw firm conclusions. It is however interesting to note that, when GW190521 is excluded, the posterior for $\Xi_0$ becomes again much closer to the result obtained with a larger prior $ m_{\rm max}\in [50, 200]\msun$ (red curve in Fig.~\ref{fig:SBRsXi0}).
In any case, care must be taken if the choice of low SNR+hard prior is made, and the impact of GW190521 must be carefully evaluated. For the purpose of this paper, the important point is that results for ${\rm SNR} > 12$ are driven by the scale $m_{\rm break}$, whose presence and position remains robust to all variations discussed so far.

\section{Comparison with the $c_M$ parametrization}\label{sec:cmparam}

Another parametrization  of modified GW propagation which  has been used in the literature is given by 
\be\label{alphaM}
\alpha_M(z)=c_M\frac{\ode(z)}{\ode}\, ,
\ee 
where 
\be
\alpha_M(z)=-2\delta(z)\, ,
\ee
and $\delta(z)$ is the function that enters in \eq{dLgwdLem}, while  $\ode(z)$ is the dark energy density fraction, and $\ode =\ode(z=0)$. This parametrization  was  first introduced  in \cite{Bellini:2014fua} to study the scalar sector of Horndeski theories and, in this case, the same function turns out to enter also in the modification of the tensor sector. In this context, it customary to use $\alpha_M(z)$ rather than $\delta(z)$.

As long as $\ode(z)$ is such that the integral in \eq{dLgwdLem} saturates to a finite value, this parametrization is still qualitatively of the same form as \eq{eq:fit}, since the ratio $\dgw(z)/\dem(z)$ goes from unity at $z=0$ to a constant $\Xi_0$ at large $z$. However,
this parametrization is less general than the $(\Xi_0,n)$ parametrization, since it only has one parameter, $c_M$, so it  implicit assumes a relation between $\Xi_0$ and $n$. 

\begin{figure}[t]
\centering
\includegraphics[width=0.45\textwidth]{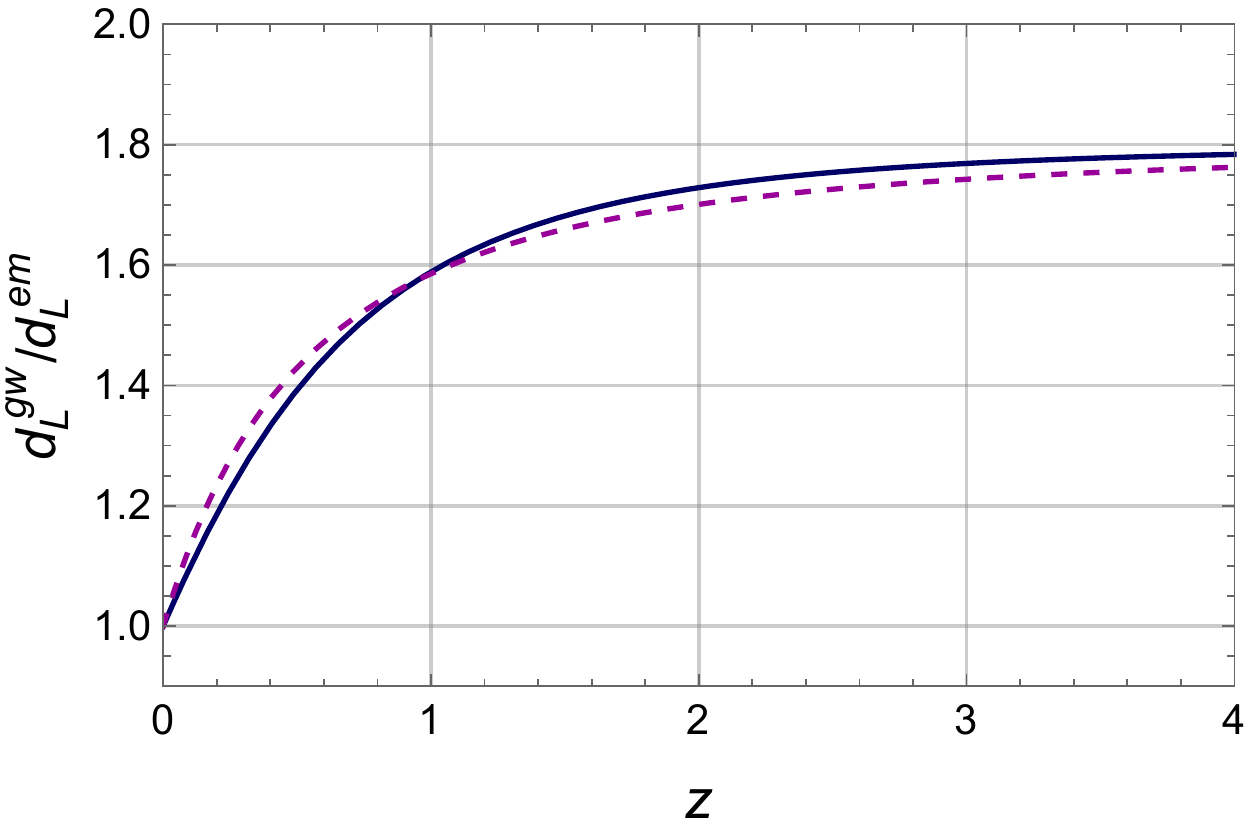}
\caption{The function $\dgw(z)/\dem(z)$ given by \eq{dgwdemcm} with $c_M=2.05$, $\oma=0.3$, $\ode=0.7$ (blue solid line) and the function $\dgw(z)/\dem(z)$ given by \eq{eq:fit} with $\Xi_0=1.8$, $n=1.9$ (magenta dashed line).
}
\label{fig:compare_cm}
\end{figure}

In order to make use, concretely, of \eq{alphaM}, one must specify  the form of $\ode(z)$. In practice, a choice that has been  done  in the literature is to assume that the dark energy density is constant in time, $\rde(z)=\rde(0)$~\cite{Lagos:2019kds} (the same choice is implicitly made in \cite{Ezquiaga:2021ayr}). Then,
denoting by $\rho_{\rm tot}(z)$ the total energy density of the Universe and by $\rho_M(z)$ the energy density in matter,  assuming a spatially flat Universe, and  observing that the energy density of radiation is irrelevant at the redshifts of interest for BBH standard sirens (even for 3G detectors),  we can write
\bees
\ode(z)&\equiv& \frac{\rde(z)}{\rho_{\rm tot}(z)}=\frac{\rde(0)}{\rde(0) +\rho_M(0) (1+z)^3}\nn\\
&=& \frac{\ode}{\ode+\oma (1+z)^3}\, ,
\ees
where $\ode=\rde(z=0)/\rho_c$, $\oma=\rho_M(0)/\rho_c$ are the present density fractions in dark energy and matter, respectively,  and $\rho_c$ is the present critical density of the Universe.
Therefore  the parametrization (\ref{alphaM}), together with the assumptions of flatness and that  the dark energy density is constant, corresponds to choosing
\be\label{deltazcM}
\delta(z)=-\frac{c_M}{2}\, \frac{1}{\ode+\oma (1+z)^3}\, ,
\ee
where $\oma+\ode=1$.
Then, carrying out the integral in \eq{dLgwdLem}, one gets~\cite{Lagos:2019kds,Ezquiaga:2021ayr}
\be\label{dgwdemcm}
\frac{\dgw(z)}{\dem(z)}=\exp\left\{ \frac{c_M}{2\ode}\log\[ \frac{1+z}{ [\oma (1+z)^3 +\ode]^{1/3} }
\]\right\}\, .
\ee
This expression has a finite limit for $z\ra\infty$ and, comparing with \eq{eq:fit},
\be\label{Xi0cm}
\log\Xi_0=\frac{c_M}{6\ode}\, \log \frac{1}{\oma}\simeq 0.29 c_M\, ,
\ee
where the last equality holds for $\oma=0.3, \ode=0.7$.
In practice, even if the functional form (\ref{dgwdemcm}) is not exactly the same as that in (\ref{eq:fit}), it is very close to it. Given $c_M$, we can determine $\Xi_0$ from \eq{Xi0cm} and adjust $n$ in \eq{eq:fit} so that the two curves are very similar. This is illustrated in Fig.~\ref{fig:compare_cm}, where the blue solid line shows the function  $\dgw(z)/\dem(z)$ from \eq{dgwdemcm}, setting $c_M=2.05$ (and $\oma=0.3$, $\ode=0.7$). This value is chosen so that \eq{Xi0cm} gives $\Xi_0=1.8$. We then adjusted $n$ to the value $n=1.9$, so that  the function $\dgw(z)/\dem(z)$ given by \eq{eq:fit} (magenta dashed line) reproduces the blue solid line quite well. Alternatively, one could fix $n$ by matching the two parametrizations near $z=0$ up to ${\cal O}(z)$, which gives 
\be
n=\frac{c_M}{2(\Xi_0-1)}\, ,
\ee 
(see also \cite{Belgacem:2019pkk,Baker:2020apq}). By construction, this choice of $n$ reproduces very well the curve
(\ref{dgwdemcm}) near $z=0$, as well as the asymptotic behavior at large $z$, but is less accurate at intermediate values of $z$.

For the purpose of phenomenological parametrizations of GW propagation, the difference between the two curves in Fig.~\ref{fig:compare_cm} is irrelevant, so  the parametrization (\ref{dgwdemcm}) is basically a subset of the $(\Xi_0,n)$ parametrization, with a built-in relation between $n$ and $\Xi_0$ (and a more complicated analytic expression for the redshift dependence). Note also that the use of \eq{alphaM} requires to specify a redshift dependence for $\ode(z)$, so the limits obtained on $c_M$ depends on this choice, while this is not the case for the $(\Xi_0,n)$ parametrization.

\section{Comparison with ref.~\cite{Ezquiaga:2021ayr}}\label{sec:comparison}

Our results for $\Xi_0$ can be compared with that found in  ref.~\cite{Ezquiaga:2021ayr}, where, as we are also doing here, a joint population-cosmology analysis is performed within a hierarchical Bayesian framework, to obtain constraints on modified GW propagation from the BBH mass function, following the original proposal in \cite{Farr:2019twy}.

Ref.~\cite{Ezquiaga:2021ayr} uses the same broken power-law mass distribution that we also use, and  the 
same Madau-Dickinson rate (\ref{Madau}). A difference with our work is that 
ref.~\cite{Ezquiaga:2021ayr}
uses only the GWTC-2 catalog~\cite{LIGOScientific:2020kqk}, that was available at the time. In particular, the analysis includes all
44 events in the GWTC-2 catalog (always including GW190521); the
corresponding detection criterion  is based on a  threshold on the FAR  which is essentially equivalent to setting a threshold on the network signal-to-noise ratio ${\rm SNR}\, > 8$.
This allows one to include more events, compared to the threshold ${\rm SNR}\, > 12$ that we use, at the price of likely introducing a non-negligible fraction of spurious events, and would also require an accurate computation of the detection probability based on the actual FAR criterion.\footnote{Note also that ref.~\cite{Ezquiaga:2021ayr} uses the LVC injections to compute selection effects. As we discussed in app.~A, this is not really correct, since these injections were performed at fixed cosmology, see the discussion below \eq{posteriorFull}, and this can introduce a bias.}
In our case, having at our disposal  the GWTC-3 catalog, and combining this with a more stringent detection threshold  ${\rm SNR}\, > 12$,  we perform the analysis on  a sample of GW events which, statistically, has a similar size (with 35 events)  but which is much more pure, and less likely to introduce biases, particularly from the computation of selection effects.

Another  significant technical difference is that the prior on $m_{\rm max}$ used in \cite{Ezquiaga:2021ayr} is 
$m_{\rm max}< 100\msun$. As we have seen in App.~\ref{sec:mmax}, when combined with a low threshold on the SNR, this has  an important effect on the result for $\Xi_0$ and, in particular, the inclusion or exclusion of GW190521 now makes a large difference; in particular, including GW190521, the posterior for $\Xi_0$ moves toward significantly lower values, while the posterior for $m_{\rm max}$ squeezes toward the upper range of the prior, see again 
Fig.~\ref{fig:GWTC3cornerXi0mmax100snr10}. This suggests that GW190521 is really an outlier of the population and, with this choice of  prior on $m_{\rm max}$, its effect is amplified;  its inclusion induces significant spurious effects on the posterior of $\Xi_0$, which is artificially moved toward lower values, to accommodate the large masses of GW190521, in a way which is very sensitive to the precise value of the upper limit chosen for the prior on $m_{\rm max}$, as long as this is chosen around $100\msun$. As we have seen in Fig.~\ref{fig:GWTC2corner}, the strong sensitivity on the upper limit of the prior on $m_{\rm max}$ basically disappears when we take it to be of order $200\msun$.

The astrophysical hyperparameters that are varied in ref.~\cite{Ezquiaga:2021ayr} are a subset of our set given in \eq{LambdaBBH}, namely $\Lambda_{\rm BBH} =\{R_0, \gamma, k, z_p,\alpha_1, \alpha_2, m_{\rm max}, b \}$. On the cosmology side, $H_0$ and $\oma$ are not varied, and  modified GW propagation
is parametrized with the $c_M$ parametrization discussed in App.~\ref{sec:cmparam}. Note also that 
ref.~\cite{Ezquiaga:2021ayr} uses 
a prior uniform in $c_M$ which,  from \eq{Xi0cm},  corresponds to a prior flat in $\log\Xi_0$.
 
The result obtained in \cite{Ezquiaga:2021ayr}, using this set of choices, 
is $c_M=-3.2^{+3.4}_{-2.0}$ (at $68\%$ C.L.). From \eq{Xi0cm}, this corresponds to  
$\Xi_0=0.4^{+0.7}_{-0.2}$. This  is  consistent, within the  large statistical uncertainty, with the results 
$\Xi_0=0.8^{+0.5}_{-0.4}$ and $\Xi_0=1.0^{+0.8}_{-0.5}$ 
that we have shown  in Figs.~\ref{fig:GWTC3cornerXi0mmax100snr12} and \ref{fig:GWTC3cornerXi0mmax100snr10},
corresponding to ${\rm SNR} > 12$ and ${\rm SNR} > 10$, respectively,
and which 
were also obtained with a prior flat in $\log\Xi_0$, setting a prior on $m_{\rm max}$ with an upper limit at $100\msun$, and including GW190521. However, as we have seen, using  a prior on $m_{\rm max}$ with an upper limit at $200\msun$, as well as the sample of the GWTC-3 catalog with a larger threshold in SNR,
${\rm SNR} > 12$,   gives more solid results, with a posterior of  $m_{\rm max}$ which is no longer squeezed on the upper limit of the prior, and the result  is no longer highly sensitive to whether we include or exclude GW190521. As a result, as we see from
Table~\ref{tab:constraintsXi0}, the limit on $\Xi_0$ moves toward higher values so that, for instance, the  $68\%$ upper limit on $\Xi_0$, that from the result of \cite{Ezquiaga:2021ayr}  would be at 
$\Xi_0\simeq 1.1$, just $10\%$ above the GR value,  becomes $\Xi_0\simeq 1.9$ including GW190521 and $\Xi_0\simeq 2.1$ excluding it. Furthermore the prior flat in $\log\Xi_0$ has the effect of further  enhancing the posterior  at low $\Xi_0$. As we see from Table~\ref{tab:constraintsXi0}, with a prior flat in $\Xi_0$ the  $68\%$ upper limit on $\Xi_0$
become $\Xi_0\simeq 2.7$ including GW190521 and $\Xi_0\simeq 2.8$ excluding it. These differences  are quite significant when one compares with the prediction $\Xi_0\simeq 1.8$ of the RT model (in the conditions that gives the highest deviation from GR, see footnote~\ref{note:inflation} on page~\pageref{note:inflation}). In the analysis of ref.~\cite{Ezquiaga:2021ayr} this would be basically excluded, while, in our analysis, this is well within the $1\sigma$ contours, see also Fig.~\ref{fig:GWTC3Xiz}.
Furthermore, as discussed above,  compared to ref.~\cite{Ezquiaga:2021ayr}, the sample of GW events that we use for our final results, shown in 
Table~\ref{tab:constraintsXi0}, has comparable size but a much higher threshold in SNR, and is therefore  much less prone to biases induced by selection effects.

For completeness,
we also observe that an analysis of the BNS GW170817 with counterpart, using the $c_M$ parametrization, was performed in~\cite{Lagos:2019kds}, and gives 
$c_M=-9^{+21}_{-28}$, corresponding to an upper bound $\Xi_0\, \lsim\, 32.5$. This can be compared to the result (\ref{delta0})  obtained earlier in \cite{Belgacem:2018lbp}, again for GW170817
(using surface brightness fluctuations rather than $H_0$ to infer the electromagnetic luminosity distance), and which corresponds to $\Xi_0\, \lsim\, 14$.

Finally, we remark that, in the context of some scalar-tensor theory of the Horndeski class, the modifications in the tensor sectors are parametrized by the same function that parametrizes the effective Newton constant  at the background level (i.e. the `running of the Planck mass'), so limits on $c_M$ can be obtained also from cosmological data of large scale structures, leading to a bound $-0.62< c_M<+1.35$ at $95\%$~c.l.~\cite{Noller:2018wyv}, which corresponds to $0.8<\Xi_0<1.7$. However, this connection is specific to these scalar-tensor  theories and does not hold in general. For instance, in the  RT non-local gravity model  \cite{Maggiore:2013mea,Belgacem:2020pdz} there is no such connection. Model-independent limits on the $(\Xi_0,n)$ parameters can be obtained only from the tensor sector.

\bibliography{myrefs}

\end{document}